\documentclass[11pt]{article}

\usepackage[T1]{fontenc}
\usepackage[utf8]{inputenc}
\usepackage{lmodern}
\usepackage{amsmath,amssymb,amsthm,mathtools}
\usepackage{hyperref}
\usepackage{geometry}
\usepackage{enumitem}
\usepackage{xcolor}

 \hypersetup{colorlinks=true,linkcolor=blue,citecolor=blue,urlcolor=blue}
\numberwithin{equation}{section}

\newcommand{\Si}[1]{{\rm Si}}
\newcommand{\Ci}[1]{{\rm Ci}}
\newcommand{\Ei}[1]{{\rm Ei}}
\newcommand{\sgn}[1]{{\rm sgn}}
\newcommand{\Span}[1]{{\rm Span}}
\newcommand{\R}{{\cal R}}
\newcommand{\A}{{\cal A}}
\newcommand{\B}{{\cal B}}
\renewcommand{\O}{{\cal O}}

\newcommand{\D}{{\cal D}}
\newcommand{\C}{{\cal C}}
\newcommand{\T}{{\cal T}}

\renewcommand{\title}[1]{\vbox{\center\LARGE{#1}}\vspace{5mm}}
\renewcommand{\author}[1]{\vbox{\center#1}\vspace{5mm}}
\newcommand{\address}[1]{\vbox{\center\footnotesize\em#1}}
\newcommand{\email}[1]{\vbox{\noindent\footnotesize\tt#1}\vspace{5mm}}

\begin{document}

\begin{titlepage}

\begin{center} 

\hfill \\
\hfill \\
\vskip 1cm

\title{Mutual Information in spacetime}

\author{Raúl E. Arias$^{a \,*}$, Marina Huerta$^{b\,\dagger}$ and Pedro J. Martinez$^{a\,\ddagger}$}

\vskip1em

\address{ \it $^{a}$ 
Instituto de Física La Plata - CONICET, \\ 
La Plata, C.C. 67, 1900, Argentina\\

\vspace{0.7em}

$^{b}$Centro Atómico Bariloche and CONICET ,\\
Instituto Balseiro, UNCuyo and CNEA,\\  S.C. de Bariloche, Río Negro, R8402AGP, Argentina \\

}
\vskip10mm

\begin{abstract}

We compute mutual information for general spacetime regions using Sorkin's Gaussian field formalism. 
We show that this construction is naturally based on the reduced space of smearings that generate independent fields, making manifest that in QFTs with stress tensor the local algebras generated by regions sharing the same causal completion. 
We test the method for a Weyl fermion, the chiral current, and massless and massive real scalar fields across various spacetime regions in 1+1 Minkowski space. Across all cases, distinct spacelike-separated spacetime regions sharing the same causal completion yield identical mutual information.
This formalism provides a setting in which to test the timelike tube theorem. Indeed, we observe that the mutual information of a timelike lens approaches that of its enveloping diamond as the cutoff is increased, offering a direct numerical verification of the theorem. In all the considered cases, our results agree with known continuum and spatial-lattice results. Finally, this work establishes a natural framework for studying entanglement measures in generalized free field theories, a task that is impractical within other existing prescriptions.

\end{abstract}

\end{center}

\vfill

\email{$^{*}$ rarias@fisica.unlp.edu.ar\\$^{\dagger}$ marina.huerta@ib.edu.ar \\$^{\ddagger}$ martinezp@fisica.unlp.edu.ar}

\end{titlepage}

\newpage

\tableofcontents

\newpage

\section{Introduction}

Information measures in quantum field theory (QFT) have become an increasingly important tool in high-energy physics; see, e.g., \cite{Casini:2009sr,Calabrese:2009qy,Nishioka:2018khk,Casini:2022rlv} for reviews. Mutual information (MI) between spacelike-separated regions in the global vacuum of the theory is a finite information measure in QFTs, known to contain details about the full spectrum of the theory and the entanglement structure of the vacuum. More recently, in the algebraic framework, MI has been formulated as a functional of the local operator algebras $\A_{\R_i}$ assigned to each region $\R_i$, $i=1,2$ \cite{Casini:2019kex, Casini:2020rgj}. Within this perspective, for example, in relativistic QFTs with a stress tensor, it is immediate to show that all operator algebras assigned to spacetime regions sharing the same causal completion are equivalent, and hence the computation of MI in any of them must yield the same result. 
While most approaches exploit this redundancy to reduce the computation to a conveniently chosen Cauchy slice,  in this work, we adopt the opposite logic: instead of reducing the algebra, we develop and test a feasible implementation of the Sorkin  formalism \cite{Sorkin2012} for algebras assigned to general spacetime regions. This implementation not only provides a natural path to verify the equivalence between algebras of geometrically distinct spacetime regions with the same causal completion but it becomes crucially useful when the reduction to a Cauchy slice is no longer valid. 

With this aim, we revisit Sorkin's formulation to compute entanglement entropy (EE) for Gaussian fields directly in terms of their spacetime correlators. Building on this, we develop both formal and numerical techniques to simplify its application to concrete examples and show how MI leads to a technically significant improvement over the EE.

We begin by considering the standard definition of the entanglement entropy of a spatial region \(R\) in a Cauchy slice \(\Sigma\) of a spacetime \(\mathcal M\). If \(\bar R\) denotes the complement of \(R\) in \(\Sigma\), and the theory is in the vacuum state \(|\Omega\rangle\), one formally defines
\begin{equation}
    S_{R}=-{\rm tr}_{R}\!\left(\rho_R \ln \rho_{R}\right),
    \qquad
    \rho_{R}\equiv {\rm tr}_{\bar R} |\Omega\rangle\langle\Omega| .
\end{equation}
Here the trace over \(\bar R\) means that we keep only the degrees of freedom assigned to \(R\). In a finite-dimensional quantum system this quantity is finite, but in continuum QFT this formula should be understood with a UV regulator. Nevertheless, while the EE in the continuum is a divergent quantity, it provides the elemental building block for MI, a more suitable measure in the continuum
\begin{equation}
I(R_1:R_2)=S(R_1)+S(R_2)-S(R_1 \cup R_2).
\label{MI-Definition}
\end{equation}
with $R_1$ and $R_2$, two non intersecting regions.
Unlike each individual entropy, this combination is finite for disconnected regions and is therefore much less sensitive to ultraviolet boundary terms. Note that according to our previous discussion when extending (\ref{MI-Definition}) to the case of $\R_1$ and $\R_2$ spacetime regions\footnote{ Throughout this work we denote subregions of a Cauchy slice with capital letters, e.g. $R$ and subregions of space-time with calligraphic capital letters, e.g. $\R$. }, then, if  $\A_{\R_1}$ and $\A_{\widetilde\R_1}$ generate the same algebra, replacing one by the other cannot change its mutual information respect to $\A_{\R_2}$ in $\R_2$. This gives a direct information-theoretic test of causal and timelike completion.
For example, for theories satisfying the time-slice property \cite{haag1962postulates}, any region containing Cauchy data for a causally complete region generates the algebra of its full domain of dependence.  
A less elementary example is provided by the timelike tube theorem \cite{Borchers,Araki}, see also \cite{Witten:2023aro,Strohmaier:2023ttt}, that demonstrates that for relativistic QFTs the algebra generated in an open region $\R$ coincides with the algebra of its timelike envelope $E(\R)$, which can be substantially larger. 
These observations will be central to the region selection in each of the cases studied here.

Several standard techniques have been developed to compute entanglement quantities; see for instance \cite{Casini:2009sr, Calabrese:2009qy, Nishioka:2018khk} for techniques that can be applied to free fields, CFTs and holography respectively. Among these tools, we distinguish  the real-time or canonical approach. 
In this method one chooses a Cauchy slice of the domain of dependence of the region and describes the theory in Hamiltonian language in terms of fields and their conjugate momenta.
For free theories, or more generally for Gaussian states, where the commutators are $c$-numbers and the state is completely determined by its two-point functions, the entropy can be obtained from the two point correlation functions restricted to the region \cite{Peschel:2002yqj}. 

Formally, it is easy to show that Sorkin's formalism is just a generalization of this approach extended to spacetime regions and reproduces the same formula for the entropy once applied to a spatial region, as discussed in more detail later in Section \ref{Sec:EE}.
The point to be  emphasized here is that the real-time approach is non manifestly covariant since it is based on a Cauchy slice. For ordinary relativistic QFTs with a Hamiltonian this is not a problem, and can even be viewed as an efficient way of using causal evolution to remove redundant spacetime information. 
In this sense, our research becomes especially relevant for theories that are Lorentz covariant but do not admit a Hamiltonian or Lagrangian description in terms of local fields. Generalized free fields provide a basic example \cite{Greenberg:1961mr}. They appear naturally, for instance, as boundary correlators of free fields in AdS \cite{Duetsch:2002hc}, and they can also be constructed intrinsically in flat spacetime, see e.g. \cite{Yngvason:1994nk}. In such theories the two-point function defines the model, while an intrinsic construction of canonical variables or of a reduced density matrix on a Cauchy slice is in general not available. Hence none of the standard methods is immediately of use and the only applicable formalism, as far as the authors know, is the covariant method discussed in this work. 

Historically, the need for a spacetime formulation of entropy appears naturally in causal set theory \cite{Bombelli:1987aa}. In that approach, spacetime is replaced by a locally finite partially ordered set, so there is no proper Cauchy slice on which to define a density matrix, while the order relation and the number of elements provide causal and volume information, respectively. 
A covariant entropy formula was proposed in this context in \cite{Sorkin2012} for Gaussian fields and explored directly in the spacetime continuum in \cite{Saravani:2013nwa}. As in \cite{Peschel:2002yqj}, the authors build on the fact that, for Gaussian states, all the information is contained in two-point functions, but rewrites the entropy in a manifestly spacetime form in terms of the Wightman function $W(x,y)$ and the Pauli-Jordan function $i\Delta(x,y)$. For a scalar field in a Gaussian state, the entropy associated with a spacetime region $\R$ is
\begin{equation}
S_\R=\sum_i\lambda_i\ln|\lambda_i|,
\label{EE-Sorkin}
\end{equation}
where the $\lambda_i$ are the generalized eigenvalues of
\begin{equation}
W\,f=\lambda\, i\Delta \,f,
\label{EE-Sorkin-2}
\end{equation}
with both kernels restricted to $\R$, and with $f$ restricted to the image of $i\Delta$. 
Both kernels $\{W,i\Delta\}$ can be represented in a finite-dimensional momentum basis, typically built out of plane wave, via a truncation procedures. We show here, that at this stage,  the generalized problem \eqref{EE-Sorkin-2}  can be solved directly projecting out the elements outside the image via a singular-value decomposition method, as explained in Section \ref{Sec:EE} .

Previous continuum applications of this prescription have mainly focused on the entropy of causally complete regions, in particular causal diamonds \cite{Saravani:2013nwa,Mathur:2021zzl} and for disconnected spacetime regions in the causal set setting in \cite{Duffy:2021disjoint}. More recently, the spectral spacetime formulation has been developed systematically for both bosonic and fermionic Gaussian theories \cite{Jones:2026spectral}.

A related, but conceptually distinct use of the Pauli-Jordan operator is the Sorkin-Johnston prescription, which defines a local vacuum state for a Gaussian theory by diagonalizing the operator $i\Delta$ restricted to a given spacetime region, and subsequently constructing the corresponding Wightman function from its positive spectral part \cite{Afshordi:2012ez,Sorkin:2011pn,Afshordi:2012jf,Johnston:2009fr}. A practical obstacle of the Sorkin-Johnston prescription is that analytic diagonalization of the Pauli-Jordan operator rapidly becomes difficult for regions with non-standard shapes or even for scalar massive free theories, see e.g. \cite{Mathur:2019yvl}. We also refer to \cite{Fewster:2013lqa} for a critique of the Sorkin-Johnston construction.

While the possible formulations of a vacuum state for a QFT defined only from its local properties is of great interest on its own, we emphasize that the entropy formula \eqref{EE-Sorkin} itself does not require this preliminary step; it requires only working restricted to the image of $i\Delta$. In our framework, the state is defined via the Wightman function, known for the global vacuum of all the theories discussed here. Only for the non-compact massless scalar will we borrow an IR regulated form of $\ W$ from the Sorkin-Johnston prescription following \cite{Saravani:2013nwa,Afshordi:2012ez}, see Section \ref{sec:massless-scalar}.

In the present work we apply this construction to a complex Weyl
fermion, the chiral current, and a real scalar field in both the
IR-regulated-massless and massive cases, in $1+1$
Minkowski spacetime. These theories have conventional canonical
descriptions and provide controlled benchmarks for the spacetime
prescription. The single-region entropies will be used only to fix the
UV truncation and to verify the numerical implementation.

For  chiral fields, the two-point functions depend on a single null coordinate and the algebra generated by a spacetime region depends only on its null projection. A causal diamond, a horizontal half-diamond, a pennant and a timelike lens with the same null projection therefore provide different spacetime representations of the same chiral algebra. We show explicitly how this equivalence appears in the generalized eigenvalue problem. We also distinguish two finite-dimensional implementations: a cutoff imposed on spacetime representatives can produce shape-dependent matrices and different convergence rates, whereas a cutoff imposed directly on the common reduced test-function space makes the matrices coincide already at finite cutoff. 
Whilst not a chiral theory on its own, we will be able to extend an analogue of this procedure to the IR-regulated massless scalar.

Finally, the massive scalar provides a qualitatively different test since the
reduction to null marginals is no longer available.
In this case, we find that the timelike tube theory holds and check our spacetime calculation against a spatial
lattice.

The paper is organized as follows. In Sec.~\ref{Sec:EE} we review the spacetime
entropy prescription, explain its implementation in a finite basis
without diagonalizing $i\Delta$, and distinguish between truncations
imposed before and after reducing the spacetime smearing space. We
also introduce the spacetime regions used to test causal and timelike
completion. In Sec.~\ref{Sec:ChiralFields} we study the Weyl fermion and the chiral
current. We derive the reduction to a single null projection and
compare shape-dependent spacetime truncations with a
quotient-adapted plane-wave truncation. In Sec.~\ref{sec:massless-scalar} we turn to the
IR massless scalar. We derive its reduction to two
compatible null marginals, construct a plane-wave basis for the
resulting quotient space, and compare the spacetime calculation with
a standard lattice computation. In Sec.~\ref{sec:massive-scalar-timelike-completion} we study the
massive scalar and test timelike completion by comparing the mutual
information of a straight-sided timelike lens with that of its
enveloping diamond. We conclude in Sec.~\ref{sec:discussion} with a discussion and future applications.

\section{Spacetime entropy and mutual information in a finite basis}\label{Sec:EE}

\subsection{Spacetime entropy for Gaussian fields}
\label{Sec:SpacetimeEntropy}

We begin by reviewing the spacetime entropy prescription for Gaussian fields proposed in \cite{Sorkin2012,Saravani:2013nwa}. Consider a bosonic operator-valued distribution $\Phi(x)$ 
in a Gaussian state $|\Omega\rangle$. The state is completely determined by its two-point function,
\begin{equation}
W(x,y)=\langle\Omega|\Phi(x)\Phi(y)|\Omega\rangle,
\label{W-def}
\end{equation}
while the commutation relations are encoded in the Pauli-Jordan kernel,
\begin{equation}
i\Delta(x,y)=[\Phi(x),\Phi(y)]
=W(x,y)-W(y,x).
\label{iDelta-def}
\end{equation}
In (\ref{W-def}) and (\ref{iDelta-def}), both kernels are defined globally by the theory and the state. A spacetime region $\cal{R}$ enters the construction by restricting their arguments and the corresponding integration domains to $\cal{R}$. The associated entropy for the boson theory is, 
\begin{equation}
S_{\rm B}(\R)=\sum_\lambda\lambda\ln|\lambda|.
\label{Entropy-Sorkin}
\end{equation}
where $\lambda$ is defined through the following generalized eigenvalue problem
\begin{equation}
\int_\R dV_y\,W(x,y)f_\lambda(y)
=
\lambda
\int_\R dV_y\,i\Delta(x,y)f_\lambda(y),
\qquad x\in \R.
\label{EV-Sorkin-Problem}
\end{equation}
The main observation of \cite{Sorkin2012} is that the eigenvalues 
$\lambda$ generalize the familiar notion of the eigenvalues of the covariance matrix for a finite collection of bosonic Gaussian variables to the continuum spacetime setting. The entropy (\ref{Entropy-Sorkin}) is then, nothing more than the von Neumann entropy of the reduced state, expressed directly in terms of these occupation modes bypassing the need to construct a reduced density matrix explicitly. For a single canonical pair, the entropy is obtained from the symplectic eigenvalues
 $\pm\sigma$, with $\sigma\geq 1/2$
\begin{equation}
S=
\left(\sigma+\frac{1}{2}\right)
\ln\left(\sigma+\frac{1}{2}\right)
-
\left(\sigma-\frac{1}{2}\right)
\ln\left(\sigma-\frac{1}{2}\right).
\label{Entropy-Sigma}
\end{equation}
It is easy to see that, for a single canonical pair $\lambda_\pm=1/2\pm\sigma$, and \eqref{Entropy-Sorkin} reduces to \eqref{Entropy-Sigma}.

 As noted in \cite{Sorkin2012}, an important point regarding (\ref{EV-Sorkin-Problem}) concerns possible null directions of the commutator. These null directions correspond to smearing functions that yield zero commutator with all observables in 
$\R$. They represent gauge or redundant degrees of freedom that do not contribute to the entropy. Equivalently, the kernel 
$i\Delta _R$ restricted to 
$\R$ need not be invertible, so \eqref{EV-Sorkin-Problem} cannot in general be replaced by an ordinary eigenvalue problem for $(i\Delta_\R)^{-1}W_\R$. The algebraically non-trivial degrees of freedom are represented on the quotient of the space of smearing functions by  $\ker \{i\Delta_\R\}$, or equivalently on the image of $i\Delta_\R$. Throughout this work, the generalized eigenvalue problem is understood on this reduced space. 
A comment is needed when the restricted commutator has null
directions. A smearing $z$ satisfying $i\Delta_R z=0$ defines an
observable that commutes with all smeared fields in $R$. For the
zero-mean Gaussian states considered here, if $W_R z=0$ as well,
this observable vanishes in the state representation and the
direction can be quotiented out. If instead $W_R z\neq 0$, the smearing $z$ contains
a fluctuating central observable and cannot be regarded
as a gauge redundancy. Thus, quotienting by the full commutator
kernel is justified only when the Wightman form also annihilates
that kernel. Otherwise, selecting a nondegenerate subspace amounts
to restricting the observable algebra. This distinction is relevant
to the finite-dimensional prescription described below.

There is an analogous construction for fermionic Gaussian fields $\Psi(x)$ \cite{Jones:2026spectral}. In this case the kernels are replaced by
\begin{equation}
W(x,y)=\langle\Omega|\Psi^\dagger(x)\Psi(y)|\Omega\rangle,\qquad\qquad C(x,y)=\{\Psi^\dagger(x),\Psi(y)\},
\label{Fermion-Anticommutator}
\end{equation}
and the generalized occupation numbers $\nu$ are obtained from
\begin{equation}
\int_\R dV_y\,W(x,y)f_\nu(y)
=
\nu
\int_\R dV_y\,C(x,y)f_\nu(y).
\label{EV-Sorkin-Fermion}
\end{equation}
For a fermionic Gaussian state the eigenvalues satisfy $\nu\in[0,1]$, and the entropy takes the form
\begin{equation}
S_{\rm F}(\R)
=
-\sum_\nu
\left[
\nu\ln\nu+(1-\nu)\ln(1-\nu)
\right],
\label{Entropy-Sorkin-Fermion}
\end{equation}
where the sum is over independent one-particle modes.

\subsection{Finite-basis implementation and UV truncation}
\label{Sec:FiniteBasis}

We now explain how the continuum generalized eigenvalue problem is implemented in practice. Let $\R$ be a bounded spacetime region and let $\{f_n(x)\}$ be a basis of smearing functions on $\R$. The restricted kernels can be represented by the matrices
\begin{equation}
\begin{aligned}
W^{\R}_{mn}
&=
\int_\R dV_x
\int_\R dV_y
f_m^*(x)W(x,y)f_n(y),
\\
(i\Delta^{\R})_{mn}
&=
\int_\R dV_x
\int_\R dV_y
f_m^*(x)i\Delta(x,y)f_n(y).
\end{aligned}
\label{Kernel-Matrix-Elements}
\end{equation}
The continuum problem \eqref{EV-Sorkin-Problem} is then replaced by
\begin{equation}
\sum_n
\left(
W^{\R}_{mn}
-
\lambda \,i\Delta^{\R}_{mn}
\right)
v_n
=
0.
\label{Discrete-Sorkin-Problem}
\end{equation}

The choice of basis is flexible and can be tailored to the specific problem at hand. One may choose basis functions that exploit the symmetries of $\R$ and the analytic properties of the kernels $W$ and $i\Delta$, which simplifies the evaluation of the integrals in \eqref{Kernel-Matrix-Elements}. Alternatively, when the null directions are annihilated by both
kernels, one may work directly in the corresponding quotient of
the smearing space and choose a basis there. This removes redundant
representatives from the outset, as illustrated in Section \ref{Sec:ChiralFields}.

For compact regions it is natural to work with a discrete family of Fourier modes. To render the continuum problem numerically tractable, we replace the infinite-dimensional space of smearing functions by a finite-dimensional subspace $\mathcal{H}_{n_{\max}}$. This is a standard UV regularization: we keep only modes with frequency below a cutoff $\omega_{\max}$.  At every value of the cutoff, \eqref{Discrete-Sorkin-Problem} is an ordinary finite-dimensional generalized eigenvalue problem. The regulated entropy is
\begin{equation}
S_{n_{\max}}(\R)
=
\sum_{\lambda}
\lambda\ln|\lambda|,\qquad\qquad \lambda\,\in\,{\rm Spec}_{\rm fin}(W^\R,i\Delta^\R)
\label{Regulated-Sorkin-Entropy}
\end{equation}
Here $\operatorname{Spec}_{\mathrm{fin}}$ denotes the generalized
spectrum on the retained nondegenerate subspace. A finite compression
can have null directions that are not null directions of the
continuum commutator. If the truncated Wightman matrix does not
annihilate such directions, discarding them is an additional
regularization choice: the resulting entropy belongs to the retained
finite observable algebra. It is not the entropy of the full
degenerate truncated algebra. We implement this choice by a
singular-value decomposition of the truncated commutator matrix,
\begin{equation}
i\Delta^\R
=
U\Sigma V^\dagger.
\label{SVD-iDelta}
\end{equation}
We retain the singular vectors associated with singular values above a fixed numerical tolerance and denote by $B$ a matrix whose columns span this numerical non-null subspace. The two kernels are then projected as
\begin{equation}
W^\R_{\rm red}
=
B^\dagger W^\R B,
\qquad
i\Delta^\R_{\rm red}
=
B^\dagger i\Delta^\R B.
\label{Reduced-Kernels}
\end{equation}
Writing $N_{\mathrm{basis}}=\dim H_{n_{\max}}$, the matrix $B$ has
dimensions $N_{\mathrm{basis}}\times r$, with orthonormal columns.
The reduced commutator is an invertible $r\times r$ matrix.

The entropy is obtained from the reduced generalized eigenvalue problem
\begin{equation}
W^\R_{\rm red}v
=
\lambda\, i\Delta^\R_{\rm red}v.
\label{Reduced-Sorkin-Problem}
\end{equation}
When the truncated commutator has full numerical rank,
$r=N_{\mathrm{basis}}$, this projection is trivial. The continuum result is defined by
\begin{equation}
S(\R)
=
\lim_{n_{\max}\rightarrow\infty}
S_{n_{\max}}(\R).
\label{Continuum-Sorkin-Entropy}
\end{equation}

A relevant comment concerns the choice of basis already at finite $n_{\rm max}$ since we will be comparing computations for several regions. Since the operator algebra assigned to a region is a notion of the continuum, one does not in general expect that two regions that define the same algebra in the continuum have the same eigenvalues (hence entropy) at finite $n_{\rm max}$. This will be the case for a general non-chiral theory and more specifically for the massive scalar we present as an example in the last section.

We now discuss the relation between $n_{\max}$ and a conventional short-distance cutoff $\epsilon$. For a region with characteristic size $l$, the large-frequency Fourier modes behave as $\omega_n\sim n/l$.
Keeping modes with $|n|\leq n_{\max}$ therefore introduces a maximal frequency
\begin{equation}
\omega_{\max}
\sim
\frac{n_{\max}}{l}.
\label{Maximum-Frequency}
\end{equation}
A spatial calculation with lattice spacing $\epsilon$ resolves frequencies of order $\omega \sim \epsilon^{-1}$. Up to order-one factors that depend on the precise choice of basis and regulator, the cutoffs are related by
\begin{equation}
\epsilon\; n_{\max}
\sim
l.
\label{Cutoff-Relation}
\end{equation}
This relation was already used in continuum applications of the spacetime prescription to compare the spectral truncation with standard CFT results \cite{Saravani:2013nwa}.

\subsection{Mutual information for spacelike separated regions}
\label{Sec:MutualInformation}

We now describe the finite-basis implementation of the MI between two disconnected spacetime regions $\R_1$ and $\R_2$. Throughout this work, the two regions will be spacelike separated. Hence, their local algebras commute, and as the usual MI defined in (\ref{MI-Definition}) for spacelike regions is finite in the continuum.

To represent the kernels (\ref{Kernel-Matrix-Elements}) on the disconnected region $\R_1\cup \R_2$, we choose independent sets of basis functions supported on $\R_1$ and $\R_2$. The truncated function space is the direct sum
\begin{equation}
\mathcal{H}_{n_{\max}}(\R_1\cup \R_2)
=
\mathcal{H}_{n_{\max}}(\R_1)
\oplus
\mathcal{H}_{n_{\max}}(\R_2).
\label{Direct-Sum-Basis}
\end{equation}
In this basis the kernel matrices take a block form,
\begin{equation}
W^{\R_1\cup \R_2}
=
\begin{pmatrix}
W_{11} & W_{12}\\
W_{21} & W_{22}
\end{pmatrix},
\qquad
i\Delta^{\R_1\cup \R_2}
=
\begin{pmatrix}
i\Delta_{11} & i\Delta_{12}\\
i\Delta_{21} & i\Delta_{22}
\end{pmatrix}.
\label{Union-Kernel-Matrices}
\end{equation}
Here, for example,
\begin{equation}
(W_{12})_{mn}
=
\int_{\R_1}dV_x
\int_{\R_2}dV_y
f^{(1)*}_m(x)
W(x,y)
f^{(2)}_n(y),
\label{Off-Diagonal-W}
\end{equation}
and the remaining blocks are defined analogously. Hermiticity of the Wightman kernel implies
\begin{equation}
W_{21}=W_{12}^\dagger.
\end{equation}

The basis of functions $f_n$ can always be chosen so that diagonal blocks $W_{11}$, $W_{22}$, $i\Delta_{11}$ and $i\Delta_{22}$ are the same matrices that enter the individual entropy problems for $\R_1$ and $\R_2$, so the only new ingredients in the union are the off-diagonal blocks, which measure correlations between the two components. Since $\R_1$ and $\R_2$ are spacelike separated, microcausality gives
\begin{equation}
i\Delta_{12}=i\Delta_{21}=0.
\label{Microcausality-Blocks}
\end{equation}
The Wightman blocks $W_{12}$ and $W_{21}$ are in general non-vanishing and encode the vacuum correlations responsible for the mutual information.

The entropy of the union is obtained by applying the same SVD reduction described in Sec.~\ref{Sec:FiniteBasis} to the block matrices \eqref{Union-Kernel-Matrices}. In the spacelike separated case the commutator matrix is block diagonal, but its individual blocks can still contain numerical null directions. We therefore project the full direct-sum space onto the numerical image of $i\Delta^{\R_1\cup \R_2}$ and solve
\begin{equation}
W^{\R_1\cup \R_2}_{\rm red}v
=
\lambda\,
i\Delta^{\R_1\cup \R_2}_{\rm red}v.
\label{Union-Generalized-Problem}
\end{equation}
The regulated mutual information is then
\begin{equation}
I_{n_{\max}}(\R_1:\R_2)
=
S_{n_{\max}}(\R_1)
+
S_{n_{\max}}(\R_2)
-
S_{n_{\max}}(\R_1\cup \R_2),
\label{Regulated-MI}
\end{equation}
where the same basis truncation and the same value of $n_{\max}$ are used in all three terms as the cancellation of ultraviolet contributions in \eqref{Regulated-MI} would in general be spoiled if the three entropies were computed with inequivalent spectral cutoffs.

The continuum mutual information is defined by
\begin{equation}
I(\R_1:\R_2)
=
\lim_{n_{\max}\rightarrow\infty}
I_{n_{\max}}(\R_1:\R_2).
\label{Continuum-MI}
\end{equation}
The cancellation of the logarithmic divergence between the entropies makes the mutual information converge considerably faster than the entropy for spacetime general regions. We stress that the construction extends immediately to regions of arbitrary shapes since its information enters only in the integration domains. When computing the MI between regions of different shapes, it will always be clear that a consistent truncation is made by keeping the same set of Fourier labels $|n|\leq n_{\max}$ on all regions.

Despite the formalism is flexible enough to allow moving the regions freely in spacetime, an important conceptual comment is that \eqref{MI-Definition} can only be properly interpreted as mutual information when the two component algebras commute \cite{Wu2025Quantum}. If two regions are timelike related, the off-diagonal commutator blocks in \eqref{Union-Kernel-Matrices} do not vanish, and the combination
\begin{equation}\label{dubious}
S(\R_1)+S(\R_2)-S(\R_1\cup \R_2) \qquad \text{ ( $\R_1$ and $\R_2$ timelike separated ) }
\end{equation}
does not in general have the standard information-theoretic properties of MI \cite{Nielsen2000Quantum}. Such configurations can be explored but will not be part of the analysis in this work. We will briefly comment on them in Sec. \ref{sec:discussion}.

\subsection{Spacetime regions and their completions}
\label{Sec:SpacetimeRegions}

We end this section by introducing the spacetime regions that will be used throughout the paper. We work in $1+1$-dimensional Minkowski spacetime and use light-cone coordinates
\begin{equation}
v=\frac{t+x}{\sqrt{2}},
\qquad
u=\frac{t-x}{\sqrt{2}}.
\label{Lightcone-Coordinates}
\end{equation}
Our reference region is the causal diamond $\mathcal{D}$ centered at the origin,
\begin{equation}
\mathcal{D}
=
\left\{
(u,v):
-l<u<l,
-l<v<l
\right\}.
\label{Diamond-Definition}
\end{equation}
This is the causal completion of the interval at $t=0$ with endpoints $x=\pm\sqrt{2}l$. The parameter $l$ will also be used to define the common Fourier scale and spectral cutoff introduced in Sec.~\ref{Sec:FiniteBasis}.

\begin{figure}[t!]
\centering
\includegraphics[width=\textwidth]{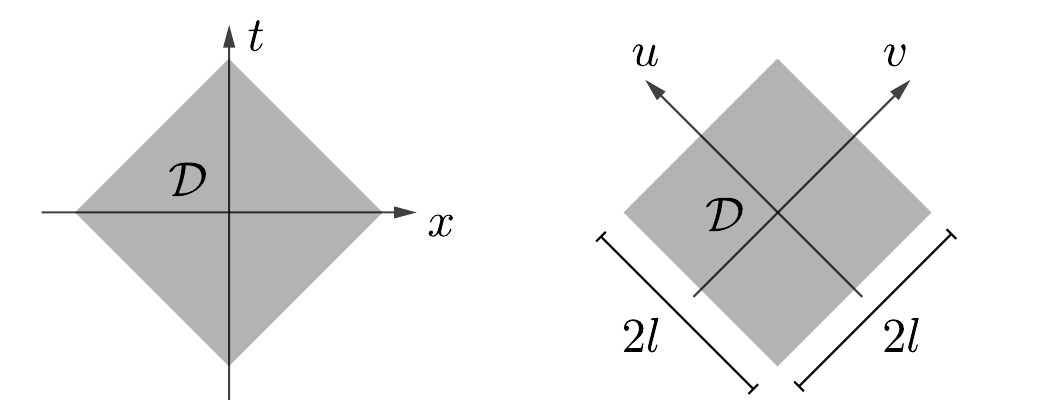}
\caption{Choice of lightcone coordinates and normalization of the spacetime diamond $\cal D$, defined by $|u|\leq l$ and $|v|\leq l$, with $v=(t+x)/\sqrt{2}$ and $u=(t-x)/\sqrt{2}$.}
\label{fig:spacetime-regions-1}
\end{figure}

\begin{figure}[t!]
\centering
\includegraphics[width=\textwidth]{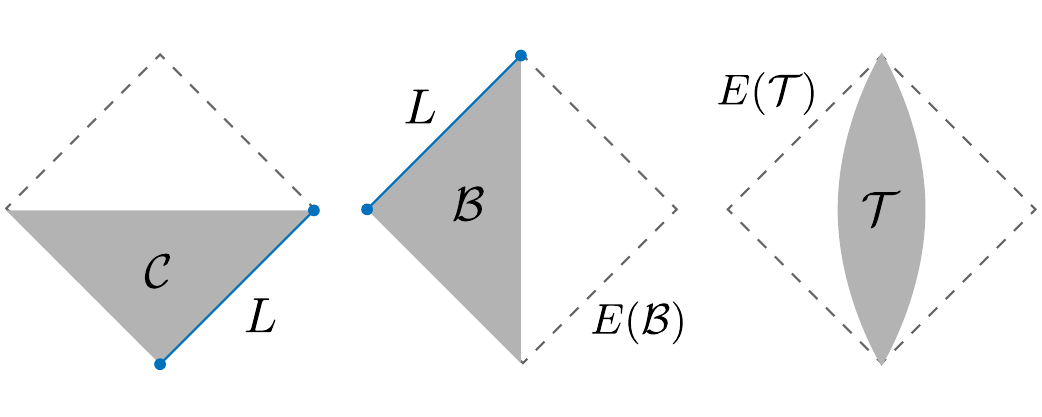}
\caption{Examples of spacetime regions contained in the diamond $\cal D$ that will be used in the EE and MI computations. The pennant $\cal B$ and lower half-diamond $\cal C$ are obtained by cutting $\cal D$ in half either in space or in time, respectively. The blue segments $L$ indicate constant-$u$ lines highlighting that in chiral theories both regions trivially build the same operator algebras. The precise constraint $u=u(v)$ defining of timelike lens $\cal T$ will take different forms in the work chosen by computation convenience. Region's $\cal C$ causal development recovers $\mathcal{D}$, whilst $\mathcal{B}$ and $\mathcal{T}$ have $\mathcal{D}$ as their timelike envelope, $E(\mathcal{B})=E(\mathcal{T})=\mathcal{D}$. }
\label{fig:spacetime-regions-2}
\end{figure}

Besides the full diamond, we will consider three regions contained in $\mathcal{D}$, shown in Fig.~\ref{fig:spacetime-regions-2}. The first one is a horizontal half-diamond, which we denote by $\mathcal{C}$. Choosing the lower half without loss of generality,
\begin{equation}
\mathcal{C}
=
\mathcal{D}\cap\{t<0\}
=
\mathcal{D}\cap\{u+v<0\}.
\label{Horizontal-Half-Diamond}
\end{equation}
The time-reflected upper half gives an equivalent configuration. The causal completion of $\mathcal{C}$ is the full diamond $\mathcal{D}$. Although the open region $\mathcal{C}$ does not contain the $t=0$ Cauchy surface of $\mathcal{D}$, it approaches it from below, and for the free fields considered below the local equations of motion imply that the algebra generated in $\mathcal{C}$ determines the algebra of the full diamond. We therefore expect
\begin{equation}
\mathcal{A}(\mathcal{C})
=
\mathcal{A}(\mathcal{D}).
\label{Time-Slice-Algebra}
\end{equation}

The second region is obtained by cutting the diamond along the timelike line $x=0$. We call either of the two resulting regions a pennant and denote it by $\mathcal{B}$. Choosing the left pennant,
\begin{equation}
\mathcal{B}
=
\mathcal{D}\cap\{x<0\}
=
\mathcal{D}\cap\{v<u\}.
\label{Pennant-Definition}
\end{equation}
The right pennant is related to it by a spatial reflection. Unlike $\mathcal{C}$, the region $\mathcal{B}$ does not contain a Cauchy surface for $\mathcal{D}$. Its relation to the full diamond is instead timelike. In the terminology of the timelike tube theorem, its timelike envelope is
\begin{equation}
E(\mathcal{B})
=
\mathcal{D}.
\label{Pennant-Envelope}
\end{equation}
Under the assumptions of the theorem, this implies
\begin{equation}
\mathcal{A}(\mathcal{B})
=
\mathcal{A}(\mathcal{D})
\label{Pennant-Algebra}
\end{equation}
for the additive algebra generated by local fields \cite{Borchers,Araki}. We stress that unlike \eqref{Time-Slice-Algebra},  eq. \eqref{Pennant-Algebra} and the timelike tube theorem is not a consequence of the causal evolution of any Cauchy data and is independent of the theory having local equations of motion.

Finally, we consider a timelike lens, denoted by $\mathcal{T}$. It is a narrow open region centered around the timelike line $x=0$, or $u=v$, and extending between the past and future tips of the diamond. A useful example region, which we use for the Weyl fermion, is
\begin{equation}
\mathcal{T}
=
\left\{
(u,v):
-l<v<l,
v-\frac{l}{2}\cos^2\left(\frac{\pi v}{2l}\right) <u<v+\frac{l}{2}\cos^2\left(\frac{\pi v}{2l}\right)
\right\}.
\label{Lens-Definition}
\end{equation}
For a massive scalar instead, we will find convenient to use another profile for $\mathcal{T}$ that we will present in due time.
Although $\mathcal{T}$ can be made arbitrarily narrow in the spatial direction, its timelike envelope is again the full diamond,
\begin{equation}
E(\mathcal{T})
=
\mathcal{D}.
\label{Lens-Envelope}
\end{equation}
The timelike tube theorem therefore predicts again
\begin{equation}
\mathcal{A}(\mathcal{T})
=
\mathcal{A}(\mathcal{D}).
\label{Lens-Algebra}
\end{equation}
The key property of $\mathcal{T}$ regardless of its shape are $\mathcal{T}\subset\mathcal{D}$ alongside eq. \eqref{Lens-Envelope}.
It is clear that the pennant and the timelike lens give two geometrically different realizations of the same statement but we find both test to be worth visiting numerically on their own. 

For chiral theories there is one further simplification. If the field depends only on the coordinate $v$, its algebra in a spacetime region depends only on the projection
\begin{equation}
\pi_v(\R)
=
\left\{
v:
(u,v)\in \R
\text{ for some }u
\right\}.
\label{Chiral-Projection}
\end{equation}
For all the regions introduced above,
\begin{equation}
\pi_v(\mathcal{D})
=
\pi_v(\mathcal{C})
=
\pi_v(\mathcal{B})
=
\pi_v(\mathcal{T})
=
L,\qquad L=\{v:v\in(-l,l)\},
\label{Equal-Null-Projections}
\end{equation}
so we expect them to represent the same chiral algebra. Notice that this statement is not related to causal or timelike completion and follows directly from chirality. Notice also that for the case of $\T$ there is no constant $u$ segment that contains $L$. In Sec.~\ref{Sec:ChiralFields} we will show how eq. \eqref{Equal-Null-Projections} appears directly in the generalized eigenvalue problem. 

The expected relations between the algebras \eqref{Time-Slice-Algebra}, \eqref{Pennant-Algebra} and \eqref{Lens-Algebra} can then be tested through mutual information. To compute mutual information, we consider two spacelike separated copies of these regions. Their associated diamonds will be denoted by $\mathcal{D}_1$ and $\mathcal{D}_2$. The comparisons are always made between regions $\R$ whose causal or timelike completions are diamonds. In this way, replacing a diamond by a horizontal half-diamond, a pennant or a timelike lens changes the spacetime integration domain while keeping fixed the local algebra that the region is expected to generate. If regions $\A(\R_{1})=\A(\widetilde{\R}_{1})$ and $\A(\R_{2})=\A(\widetilde{\R}_{2})$, then one must have
\begin{equation}
I(\R_1:\R_2)=I(\widetilde{\R}_1:\R_2)=I(\R_1:\widetilde{\R}_2)=I(\widetilde{\R}_1:\widetilde{\R}_2).
\label{Algebra-MI-Equality}
\end{equation}
In particular, eqs. \eqref{Time-Slice-Algebra}, \eqref{Pennant-Algebra}, \eqref{Lens-Algebra} and \eqref{Algebra-MI-Equality} imply
\begin{equation}
I(\mathcal{X}_1:\mathcal{Y}_2)
=
I(\mathcal{D}_1:\mathcal{D}_2),
\qquad
\mathcal{X},\mathcal{Y}
\in
\left\{
\mathcal{D},
\mathcal{C},
\mathcal{B},
\mathcal{T}
\right\},
\label{Completion-MI-Prediction}
\end{equation}
For chiral theories, the same prediction follows already from the equality of the null projections. The rest of the paper will examine in several examples how these equivalences are realized by the spacetime generalized eigenvalue problem.

We should clarify that we use a convention such that whenever we discuss configurations of the form, e.g. $(\mathcal B_1,\mathcal B_2)$ it is left implicit that $\B_1$ is a left pennant and $\B_2$ is a right pennant. The same goes for $\C_1$ and $\C_2$ that will be respectively be lower half and upper half diamonds respectively. This is not of much importance for the algebras as to obtain faster numeric convergence, since it often retains nice symmetries in the integrands.

We close this section by mentioning that we will focus in studying the mutual information as a function of the cross-ratio defined by our regions $\R_1$ and $\R_2$. For two disconnected segments at $t=0$ (whose causal completion result in the diamonds $\D_1$ and $\D_2$ ) with ordered endpoints $x_1<x_2<x_3<x_4$, we define the conformal cross-ratio
\begin{equation}
\eta
=
\frac{(x_2-x_1)(x_4-x_3)}
{(x_3-x_1)(x_4-x_2)}.
\label{Cross-Ratio}
\end{equation}
The limit $\eta\rightarrow0$ corresponds to large separation, while $\eta\rightarrow1$ corresponds to the two diamonds approaching each other. 
For conformal theories in 1+1 dimensions as will be the case for our examples in Sec. \ref{Sec:ChiralFields}, it can be shown that the mutual information is only a function of the cross-ration $\eta$. For the non-conformal examples, the mutual information also depends
on dimensionless combinations involving the physical scales.
For equal diamonds of size $l$, these are $ml$ for the massive
scalar and $\mu l$ for the infrared-regulated massless scalar.
We therefore study the dependence on $\eta$ at fixed $ml$ or
$\mu l$, respectively. At each value of $\eta$, comparisons between
different spacetime realizations use the same associated diamonds
and the same state parameters.

\section{Chiral fields}
\label{Sec:ChiralFields}

We begin our applications with chiral fields. These theories provide the cleanest setting in which to separate the geometry of a spacetime region from the operator algebra that it generates. We consider $v$-chiral fields, which depend only on the null
coordinate $v$ and whose local algebras are naturally assigned
to intervals of the chiral line \cite{CasiniHuerta2009,Arias:2018}. The Weyl fermion and the chiral current studied below have different statistics and short-distance singularities, but the relation between spacetime regions and their algebras follows from the same kinematical argument.

\subsection{Chiral algebras from spacetime regions}
\label{Sec:ChiralProjection}

From the operator-algebraic point of view, a chiral conformal field
theory is naturally described by a net of local algebras associated
with intervals of a light ray 
\cite{Rehren:1999chiral,Rehren:2015acqft,Kawahigashi:2017cft}.
In particular, the algebra of left- or right-chiral observables
localized in a two-dimensional double cone depends only on the
corresponding light-ray interval \cite{Rehren:1999chiral}. In the
present spacetime formulation, this familiar statement is realized
explicitly by projecting a two-dimensional smearing function onto the
null coordinate on which the chiral field depends. Let $\O(v)$ be a chiral operator-valued distribution and let $\R$ be an open spacetime region in the $(u,v)$ plane. A smeared operator supported in $\R$ takes the form
\begin{equation}
\O_\R[f]
=
\int_\R du\,dv\,f(u,v)\O(v).
\label{Chiral-Smeared-Operator}
\end{equation}
Since the field is independent of $u$, this can be rewritten as
\begin{equation}
\O_\R[f]
=
\int_{\pi_v(\R)}dv\,F_\R(v)\O(v),
\qquad
F_\R(v)
=
\int_{\R_v}du\,f(u,v),
\label{Effective-Chiral-Smearing}
\end{equation}
where
\begin{equation}
\R_v
=
\left\{
u:(u,v)\in \R
\right\}
\label{Region-Fiber}
\end{equation}
is the section of the region at fixed $v$.
A useful surjective map can be built as follows. Let 
\begin{equation}
T_\R:C^\infty_c(\R)\longrightarrow C^\infty_c(\pi_v(\R)),
\qquad
(T_\R f)(v)=\int_{\R_v}du\,f(u,v).
\label{Chiral-Projection-Map}
\end{equation} 
For $F\in C^\infty_c(\pi_v(\R))$, 
one can always choose a smooth function
$\chi_\R(u,v)$, supported inside $\R$, such that
\begin{equation}
\int_{\R_v}du\,\chi_\R(u,v)=1
\label{Normalized-Transverse-Smearing}
\end{equation}
on the support of $F$. The spacetime smearing
\begin{equation}
f_F(u,v)=F(v)\chi_\R(u,v)
\label{Lifted-Chiral-Smearing}
\end{equation}
then satisfies $T_\R f_F=F$.

For example, for the pennant \eqref{Pennant-Definition}, 
\begin{equation}
\mathcal B
=
\{(u,v):-l<v<l,\ v<u<l\},
\end{equation}
one may take 
\begin{equation}
\chi_{\mathcal B}(u,v)
=
\frac{1}{l-v}
\chi\left(\frac{u-v}{l-v}\right),
\qquad
\chi\in C^\infty_c(0,1),
\qquad
\int_0^1ds\,\chi(s)=1.
\label{Pennant-Normalized-Lift}
\end{equation}
For every $F$ with compact support in $(-l,l)$ this gives
\begin{equation}
\int_v^l du\,F(v)\chi_{\mathcal B}(u,v)=F(v).
\end{equation}

Consequently, the space of spacetime smearings modulo the directions
which are invisible to the chiral field is naturally isomorphic to the
test-function space on the null projection,
\begin{equation}
\frac{C^\infty_c(\R)}{\ker T_\R}
\simeq
C^\infty_c(\pi_v(\R)).
\label{Chiral-Smearing-Quotient}
\end{equation}
In particular,
\begin{equation}
\mathcal A_\O(\R)
=
\mathcal A_\O\bigl(\pi_v(\R)\bigr).
\label{Chiral-Algebra-Projection}
\end{equation}
Thus two spacetime regions with the same projection onto the chiral
null coordinate generate the same algebra.
For the regions introduced in Sec.~\ref{Sec:SpacetimeRegions}, one therefore has
\begin{equation}
\mathcal{A}_\O(\mathcal{D})
=
\mathcal{A}_\O(\mathcal{C})
=
\mathcal{A}_\O(\mathcal{B})
=
\mathcal{A}_\O(\mathcal{T}).
\label{Equal-Chiral-Algebras}
\end{equation}

To illustrate the distinction between a spacetime representative and
the intrinsic projected algebra, consider the particular
choice of spacetime smearings which are constant along the transverse
null direction,
\begin{equation}
f_n^\R(u,v)=h_n(v),
\label{Constant-Transverse-Chiral-Smearing}
\end{equation}
and let for each $v$ define the width $w_\R(v)\in\mathbb{R}$ in the $u$ coordinate as
\begin{equation}
\int_{\R_v} du = w_\R(v)\,.
\end{equation}

Now, for a chiral kernel $K(v,v')$ independent of $u$ and $u'$, the
matrix elements obtained from these representatives are
\begin{equation}
K^\R_{mn}
=
\int_{\pi_v(\R)}dv
\int_{\pi_v(\R)}dv'\,
 \left(w_\R(v) h_m(v)\right)^*
K(v,v')
\left(h_n(v')w_\R(v')\right).
\label{Chiral-Region-Matrix}
\end{equation}
where we see that our choice of basis in \eqref{Constant-Transverse-Chiral-Smearing} can equivalently be viewed as
the matrix elements of the interval kernel in the weighted projected functions
\begin{equation}
h^\R_n(v)=T_\R h_n(v)=w_\R(v)h_n(v).
\label{Weighted-Chiral-Basis}
\end{equation}
A formal transverse-constant representative with prescribed
projection is
\begin{equation}
g_n^\R\equiv\frac {h_n(v)}{w_\R(v)}\qquad\Rightarrow\qquad T_\R g_n^\R=h_n(v).
\label{Quotient-Adapted-Lifts}
\end{equation}
This representative need not belong to $C_c^\infty(\mathcal R)$
or to $L^2(\mathcal R)$ when the width vanishes at the boundary.
For smooth compactly supported projected functions, genuine smooth
lifts are instead provided by the construction in
\eqref{Normalized-Transverse-Smearing}-\eqref{Lifted-Chiral-Smearing}. More generally, the projected functions used in
a finite-basis calculation must belong to the domain of the
relevant correlation forms. The matrix identities below depend
only on their projected data, independently of the particular
choice of spacetime representatives.
The advantage of this can be seen directly at
the level of the kernel matrices. Let $\R$ and $\widetilde\R$ be two different
spacetime regions, with the same null projections
\begin{equation}
L=\pi_v(\R)=\pi_v(\widetilde\R),
\end{equation}
and let $g_m^{\R}$ and $g_m^{\widetilde\R}$ be lifts of
the same projected functions $h_m$, so that
\begin{equation}
T_{\R}g_m^{\R}=T_{\widetilde\R}g_m^{\widetilde\R}=h_m,
\label{Quotient-Adapted-Lift-Conditions}
\end{equation}
For any chiral kernel $K(v,v')$, it follows that 
\begin{align}
K_{mn}^{\R}=
\int_{L}dv
\int_{L}dv'\,
h_m(v)^*
K(v,v')
h_n(v')=K_{mn}^{\widetilde\R}.
\label{Quotient-Adapted-Kernel-Reduction}
\end{align}
entry by entry. 
It is straightforward to extend this argument to the matrix elements involved in the computation of mutual information which involves two regions $\R_1$ and $\R_2$ at spacelike distance.
Applying this identity to both kernels entering the
bosonic or fermionic generalized eigenvalue problem shows that the
complete matrix, including their local and crossed blocks,
coincide at every finite $n_{\max}$. Hence, they generate the same algebra in the continuum and their EE and MI also coincide trivially.

The two choices $h^\R_n(v)$ and $g^\R_n(v)$ generate the same continuum chiral algebra, but define
different finite-dimensional approximations to it. The functions
$h^\R_n=w_\R\,h_n$ define a ``shape-weighted'' spacetime truncation, whereas
\eqref{Quotient-Adapted-Lifts} defines a truncation directly adapted
to the quotient test-function space on the null projection.
We use the shape-weighted truncation for
the Weyl fermion, where it provides a useful test of regulator
robustness, and the quotient-adapted truncation for the chiral current,
whose more singular two-point function makes the shape-weighted
convergence considerably less stable.

\subsection{Weyl fermion}
\label{Sec:WeylFermion}

The complex Weyl fermion $\psi(v)$ provides the simplest example in which to apply the general discussion of Sec.~\ref{Sec:ChiralProjection}. We choose a $v$-chiral field and work in the Poincar\'e-invariant
vacuum. The Wightman and anticommutator kernels are
\begin{equation}
\begin{aligned}
W_\psi(v,v')
&=
\langle\Omega|\psi(v)\psi^\dagger(v')|\Omega\rangle
=
\frac{1}{2\pi i}
\frac{1}{v-v'-i0^+},
\\
C_\psi(v,v')
&=
\left\{
\psi(v),\psi^\dagger(v')
\right\}
=
\delta(v-v').    
\end{aligned}
\label{eq:psi-W-C}
\end{equation}
Both kernels depend only on the chiral coordinate. Therefore, for any spacetime region $\R$, their matrix elements are completely determined by its null projection and by the effective width $w_\R(v)$ for the regions presented in Fig. \ref{fig:spacetime-regions-2}, with $\T$ defined by \eqref{Lens-Definition}. 
The entropy is computed from the generalized eigenvalues in \eqref{EV-Sorkin-Fermion} and from \eqref{Entropy-Sorkin-Fermion}.

For a single interval of length $L_V$, the continuum entropy is known to be
\begin{equation}
S_\psi(V)
=
\frac{1}{6}
\ln\left(\frac{L_V}{\epsilon}\right)
+
c_\psi,
\label{psi-EE-an-result}
\end{equation}
where $c_\psi$ is regulator dependent. Since all the spacetime regions considered in Sec.~\ref{Sec:SpacetimeRegions} project onto the interval $L$ of legth $2l$, they have the same continuum chiral entropy. In terms of the spectral truncation, the universal statement is
\begin{equation}
S_{\psi,n_{\max}}(\R)
=
\frac{1}{6}\ln n_{\max}
+
c_\R
+
o(1),
\qquad
\R\in
\left\{
\mathcal D,
\mathcal C,
\mathcal B,
\mathcal T
\right\}.
\label{psi-EE-kmax}
\end{equation}
The constants and the rate of convergence can depend on the spacetime representation of the algebra. In particular, the diamond EE converges considerably faster than regions whose width functions mix a larger number of Fourier modes. The logarithmic coefficient $1/6$ is nevertheless recovered for all the regions considered.

\begin{figure}[t]
\centering
 \makebox[\textwidth][c]{\includegraphics[width=1.12\textwidth]{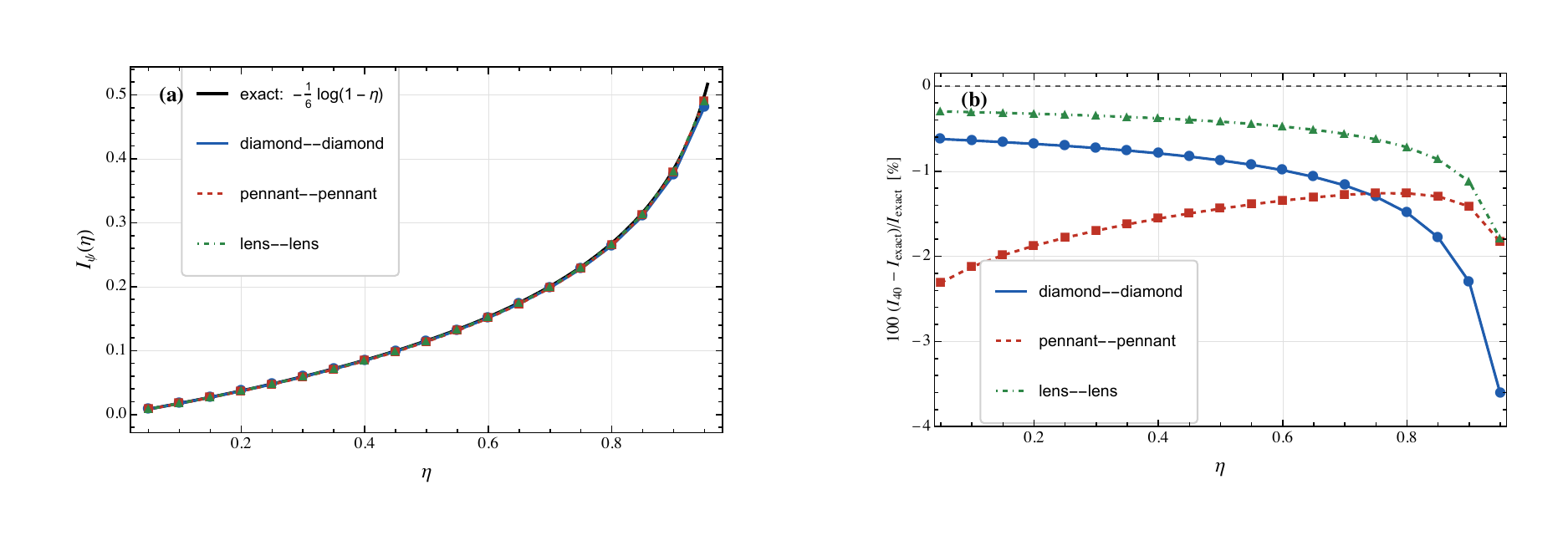}}
\caption{Mutual information of the complex Weyl fermion for different spacetime representations of the same pair of chiral algebras. Panel (a) shows the finite-basis Sorkin results at $n_{\max}=40$ for two causal diamonds, two pennants and two timelike lenses, together with the exact continuum result $I_\psi^{\mathrm{exact}}(\eta)
=
-\frac{1}{6}\ln(1-\eta)$.
Panel (b) shows the corresponding percentage deviations, 
for ${\cal X}{\cal Y}=\mathcal{D}\mathcal{D}$, $\mathcal{B}\mathcal{B}$ and $\mathcal{T}\mathcal{T}$. Although the different spacetime regions produce different truncated kernel matrices and different finite-cutoff errors, their mutual informations remain close to the same exact function determined by their common null projections. Notice that the $\mathcal{T}\mathcal{T}$ configuration beats the $\mathcal{D}\mathcal{D}$ configuration consistently for all $\eta$.}
\label{fig:Weyl-MI}
\end{figure}

For the Weyl fermion the mutual information is known exactly,
\begin{equation}
I_\psi^{\mathrm{exact}}(\eta)
=
-\frac{1}{6}\ln(1-\eta),
\label{Weyl-Exact-MI}
\end{equation}
where $\eta\in(0,1)$ 
is the cross-ratio defined in eq. 
\eqref{Cross-Ratio}. 
We evaluate the Sorkin mutual information for the representative configurations
\begin{equation}
\mathcal D\mathcal D=(\mathcal D_1,\mathcal D_2),
\qquad
\mathcal C\mathcal C=(\mathcal C_1,\mathcal C_2),
\qquad
\mathcal B\mathcal B=(\mathcal B_1,\mathcal B_2),
\qquad
\mathcal T\mathcal T=(\mathcal T_1,\mathcal T_2).
\label{Weyl-Configurations}
\end{equation}
In every case, the associated null projections are the same pair of intervals. With the orientations specified in Sec.\ref{Sec:SpacetimeRegions}, the corresponding
half-diamonds and pennants also have identical width functions,
$w_{\mathcal C_i}=w_{\mathcal B_i}$ for $i=1,2$.
Their truncated kernel matrices therefore coincide, so the
$\C\C$ and $\B\B$ mutual informations agree at every cutoff.
Only the $\B\B$ curve is shown in Fig.\ref{fig:Weyl-MI}.

We represent the kernels in the Fourier basis
\begin{equation}
h_n(v)
=
e^{-i\pi nv/l},
\qquad
n\in\mathbb{Z},
\qquad
|n|\leq n_{\max}.
\label{eq:psi-basis}
\end{equation}
At finite $n_{\max}$, the results obtained from the different spacetime regions need not coincide exactly, since their width functions lead to different truncated kernel matrices. Nevertheless, already at $n_{\max}=40$ all the configurations considered reproduce the exact continuum result within a few percent over the full range of cross-ratios shown. The remaining discrepancies become more visible as $\eta\to1$, where increasingly short-distance modes contribute to the mutual information, and will be interpreted as finite-truncation effects. Figure~\ref{fig:Weyl-MI} compares the different spacetime configurations at fixed cutoff. 

An interesting result of Fig.~\ref{fig:Weyl-MI} is that it shows that the $\mathcal{T}\mathcal{T}$ configuration beats the $\mathcal{D}\mathcal{D}$ configuration consistently for all $\eta$ in the mutual information, despite converging slower to the UV-divergent continuum limit of the entanglement entropy. This is perhaps unexpected in terms of symmetry since $\mathcal{T}$ mixes the $u$ and $v$ domains in a highly non-trivial way, leading to more convoluted integrals, but it is fairly reasonable from a physical point of view as we now explain\footnote{We are indebted to Christopher J. Fewster for the explanation of this result.\label{Fewster}}. For our current example of a massless Weyl fermion, our theory is conformal so that the modular flow inside $\D$ is realized locally by a conformal map from the Rindler case and its flow can be identified with Rindler observers. One can notice that the observers ``missing'' in $\mathcal{T}$ with respect to $\mathcal{D}$ are the ones that are highly accelerated and that should actually need a higher cutoff to be resolved. 
Hence, we find that the diamond is inefficiently trying to incorporate high energy operators in the algebra to reproduce the corners at $t=0$ which are very expensive to obtain at any value in the truncated finite $n_{\max}$ algebra. On the contrary, the smooth boundaries of $\mathcal{T}$ remove this highly energetic operators and for the same value of $n_{\max}$ it has already incorporated more relevant operators in the algebra that draw the result closer to \eqref{Weyl-Exact-MI}. This physical effect cannot be distinguished by simply comparing entanglement entropy for which $\D$ appears to converge faster to the correct value of the central charge $c=1/6$, but we argue that EE is not itself a proper quantity in the UV theory.

It is reasonable to expect similar results to hold in general relativistic theories and regions, since the modular flow always approximates the local Rindler result sufficiently close to the boundary in space, i.e. one should expect a smooth timelike lens $\mathcal{T}$ to provide an equal or better convergence rate than a spacetime diamond $\mathcal{D}$ when compared using the same basis and at the same value of the cut-off. The trade-off is that diamond matrix elements are usually easier to compute, but we find this result still physically relevant on its own. In particular, we will not work with a smoothed sided $\mathcal{T}$ for the massive scalar in Sec.~\ref{sec:massive-scalar-timelike-completion} and find that the diamond in that case converges faster than the straight-edged timelike lens.

The Weyl fermion therefore realizes the null-projection equivalence in its simplest form. For our choice of basis, the spacetime matrices depend on the shape of the region and can have substantially different finite-cutoff behavior, but their continuum mutual information is fixed entirely by the projected intervals. Had we instead truncated directly in the
quotient space $C^\infty_c(\R)/\ker T_\R\simeq C^\infty_c(\pi_v(\R))$, using the same
projected basis for every spacetime realization, the corresponding
finite matrix pencils would coincide at every value of $n_{\max}$. In the next subsection we follow this approach for the chiral current.

\subsection{Chiral current}
\label{Sec:ChiralCurrent}

We now apply the null-projection construction to the chiral current. We define $j(v)$ 
with vacuum kernels
\begin{equation}
W_j(v,v')
=
-\frac{1}{4\pi}
\frac{1}{(v-v'-i0^+)^2},
\qquad
i\Delta_j(v,v')
=
\frac{i}{2}\delta'(v-v').
\label{Current-Kernels}
\end{equation}

\begin{figure}[t!]
\centering
\includegraphics[width=0.78\textwidth]
{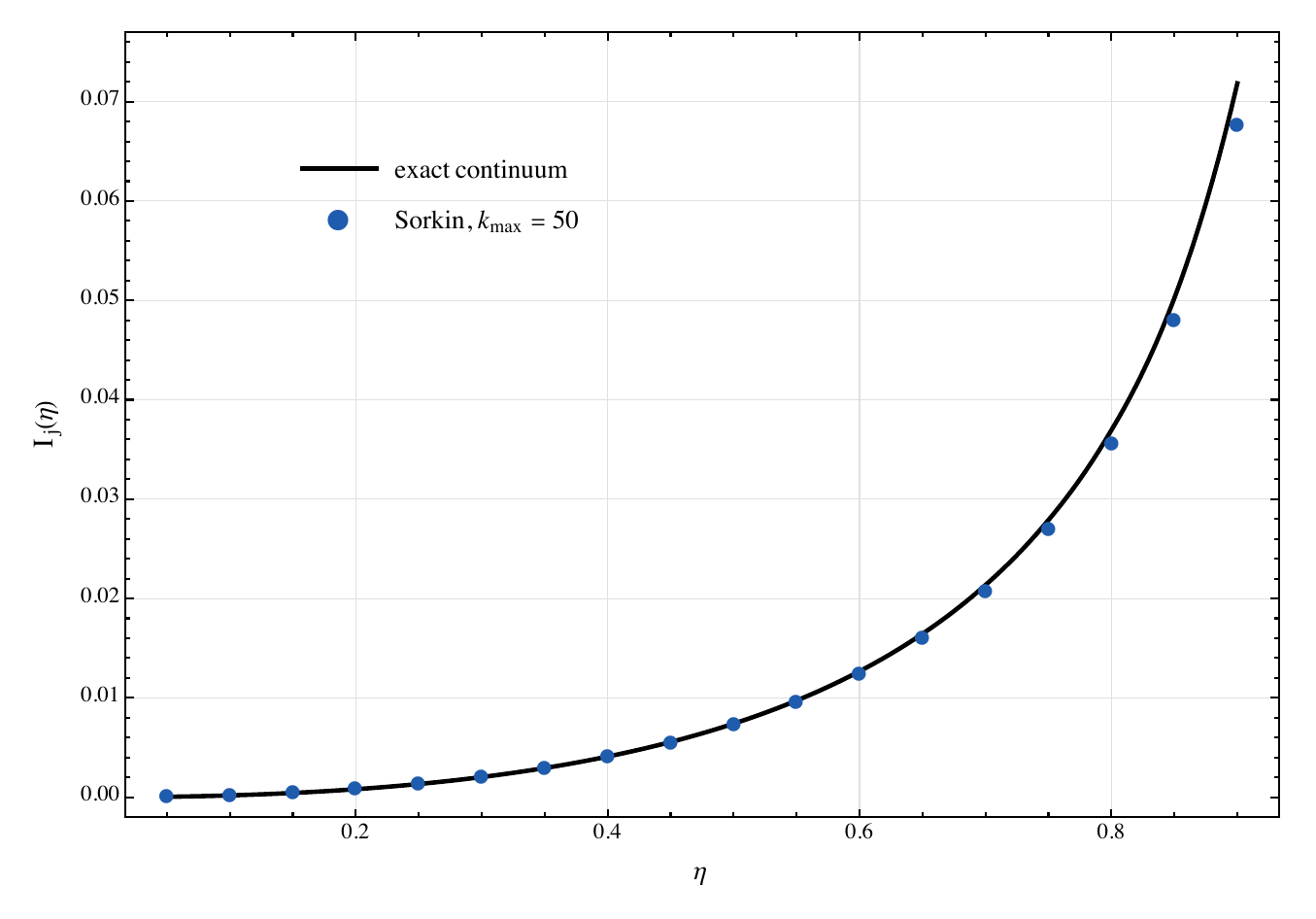}
\caption{Mutual information of the additive chiral-current algebra.
The points are obtained from the Sorkin generalized eigenvalue problem
in the quotient-adapted plane-wave basis at $n_{\max}=50$. The solid black
curve is the exact continuum result
\eqref{Current-Exact-MI} of \cite{Arias:2018}. In our basis the same finite-cutoff result applies to all the spacetime pairs listed in
\eqref{Completion-MI-Prediction}.}
\label{fig:Current-MI}
\end{figure}

For every spacetime realization of the same projected interval $L$, we define $g^\R_n$ that truncate to the same $h_n$ defined in \eqref{eq:psi-basis} after passing to the projected algebra. We have already shown that the $g^\R_n$ basis ensures that matrices for all regions are the same as the ones obtained by integrating only over $L$. Thus, we only perform this numerical test for the diamond in the basis \eqref{eq:psi-basis} against analytical results known from the literature.

Two comments are of interest regarding the matrix elements computation for the chiral current. First, we note that naive double integration of $W_j(v,0)\sim v^{-2}$ against our basis \eqref{eq:psi-basis} leads to log divergences as the UV cutoff distance is removed. This is just a reminder that QFT kernels are defined as distribution-valued functions, so the proper definition of this matrix integrals follow from moving the derivatives onto the projected smearing functions before
evaluating the resulting Fourier integrals. The second comment concerns the $n=0$ mode of our basis, which easily seen to be in $\ker \{i\Delta_j\}$. We have checked that this is the only element of our basis in $\ker \{i\Delta_j\}$ and that removing the element by hand and removing it after computing the full matrices via the SVD described in Sec.~\ref{Sec:FiniteBasis} yields identical results. This is an important test for the robustness of the method, considering the possibility that for more involved $i\Delta$ operators, multiple non-trivial combinations of a given basis may unexpectedly belong to $\ker \{i\Delta\}$ as will be the case for the massless scalar.

The continuum mutual information of the additive current algebra is
known analytically from Ref.~\cite{Arias:2018}. In our conventions it can be
written as
\begin{equation}
I_{j}(\eta)
=
-\frac{1}{6}\log(1-\eta)
+
U(\eta),
\label{Current-Exact-MI}
\end{equation}
where
\begin{equation}
U(\eta)
=
\pi
\int_0^\infty ds\,
\frac{s}{\sinh^2(\pi s)}
\operatorname{Im}\left[\,
\ln\, _2F_1(1+is,-is;1;\eta)\,\right].
\label{Current-Arias-U}
\end{equation}

Figure~\ref{fig:Current-MI} compares this continuum result with the
Sorkin generalized eigenvalue problem at $n_{\max}=50$. The numerical
curve follows the exact result over the full range shown. 
The discrepancy 
increases as the intervals approach each
other, as expected from a UV truncation.

\section{The IR-regulated massless scalar}
\label{sec:massless-scalar}

\subsection{Null reduction and plane-wave truncation}
\label{sec:scalar-null-reduction}

We now turn to the real non-compact massless scalar in
$1+1$ dimensions. Its infrared subtlety is that the standard
Poincar\'e-invariant vacuum two-point function is not defined on
arbitrary test functions. Derivative observables, or scalar
smearings with vanishing total integral, avoid this obstruction.
To include the full scalar smearing space, an infrared
prescription for the state is required. To provide a vacuum state for our IR regulated theory, we choose to follow \cite{Afshordi:2012ez,Saravani:2013nwa} and work with the SJ vacuum \cite{Sorkin:2011pn,Johnston:2009fr} for a massless scalar theory confined to a finite spacetime diamond of size of order $\mu^{-1}$. This is a consistent way of regulating the theory and provides a unique vacuum state correlator. For our purposes, we will not need the full form of the correlator but consider the approximation in which we want to compute EE and MI of systems of characteristic length $l$ much smaller than the $\mu^{-1}$ IR cutoff. In considering two regions, they should also not be taken too far away as to compete with the scale $\mu^{-1}$. Thus, our computations hold as long as all relevant distances are smaller than $\mu^{-1}$.

In this local regime, we use the leading logarithmic approximation
to the SJ two-point function, i.e.
\begin{equation}
W_{\phi,\mu}(x,x')
=
-\frac{1}{4\pi}
\ln\!\left[
-\mu^2
(u-u'-i0^+)(v-v'-i0^+)
\right].
\label{Scalar-Wightman}
\end{equation}
since it specifies the infrared prescription for the state\footnote{ It is interesting to point out that carrying out MI calculations with the $\mu$ prescribed by the SJ vacuum matches exactly with the result coming from a self-dual radius compactification of the compact scalar. One could in principle move $\mu$ to find a phenomenological relation between these regulators.}, but for our purposes it is enough that we make sure that $\mu^{-1}$ is the largest length scale in our computations and that it must be kept
fixed when comparing different spacetime regions since it defines the vacuum and, hence, the theory.
Its antisymmetric part is independent of $\mu$,
\begin{equation}
i\Delta_\phi(x,x')
=
-\frac{i}{4}
\left[
\operatorname{sgn}(u-u')
+
\operatorname{sgn}(v-v')
\right].
\label{Scalar-Pauli-Jordan}
\end{equation}

We now develop for this theory an extension of the null-projection technique we introduced for chiral theories in the previous section. We begin by noticing that the equation of motion is
\begin{equation}
\partial_u\partial_v\phi(u,v)=0,
\end{equation}
and therefore its local solutions can be written as
\begin{equation}
\phi(u,v)=\phi_u(u)+\phi_v(v).
\label{Scalar-Chiral-Decomposition}
\end{equation}
However, the two terms on the right-hand side should
not be interpreted as two completely Wightman fields, for they
share a zero-mode constraint, which can be seen as the origin of the
infrared subtlety of the massless scalar in two dimensions. The
description in terms of two chiral sectors with a common zero mode has been already noticed in the literature, see for example
\cite{Derezinski:2004dm,Ciolli:2005ci}. Here we only use
\eqref{Scalar-Chiral-Decomposition} as a statement about the dependence
of smeared solutions on their null coordinates.

Let $\R$ be any of the open spacetime regions introduced in
Sec.~\ref{Sec:SpacetimeRegions}, and let $f\in C_c^\infty(\R)$. In
addition to the sections $\R_v$ defined in \eqref{Region-Fiber}, we
introduce
\begin{equation}
\R_u
=
\left\{
v:(u,v)\in \R
\right\}.
\end{equation}
The two null marginals of $f$ are then
\begin{equation}
G_f(u)
=
\int_{\R_u}dv\,f(u,v),
\qquad
F_f(v)
=
\int_{\R_v}du\,f(u,v).
\label{Scalar-Null-Marginals}
\end{equation}
Using \eqref{Scalar-Chiral-Decomposition}, the smeared scalar can be
written as
\begin{equation}
\phi_\R[f]
=
\int_{\pi_u(\R)}du\,G_f(u)\phi_u(u)
+
\int_{\pi_v(\R)}dv\,F_f(v)\phi_v(v).
\label{Scalar-Smeared-Reduction}
\end{equation}
Thus the smeared field is insensitive to the part of $f$ which is not contained
in the pair $(G_f,F_f)$. This defines the map
\begin{equation}
T_\R:C_c^\infty(\R)
\longrightarrow
C_c^\infty\bigl(\pi_u(\R)\bigr)
\oplus
C_c^\infty\bigl(\pi_v(\R)\bigr),
\qquad
T_\R f=(G_f,F_f).
\label{Scalar-Null-Reduction-Map}
\end{equation}
The crucial point is that the two components in the image of $T_\R$ are not independent, since
\begin{equation}
\int_{\pi_u(\R)}du\,G_f(u)
=
\int_\R du\,dv\,f(u,v)
=
\int_{\pi_v(\R)}dv\,F_f(v).
\label{Scalar-Compatibility}
\end{equation}
Therefore the compatible null-data space is actually
\begin{equation}
\mathcal Q_\R
=
\left\{
(G,F)\in
C_c^\infty\bigl(\pi_u(\R)\bigr)
\oplus
C_c^\infty\bigl(\pi_v(\R)\bigr):
\int du\,G(u)=\int dv\,F(v)
\right\}.
\label{Scalar-Quotient-Space}
\end{equation}
The condition in \eqref{Scalar-Quotient-Space} also removes the
ambiguity in the decomposition \eqref{Scalar-Chiral-Decomposition}, i.e. the transformation
\begin{equation}
\phi_u\longrightarrow\phi_u+c,
\qquad
\phi_v\longrightarrow\phi_v-c
\end{equation}
does not change \eqref{Scalar-Smeared-Reduction} provided that the two
integrals in \eqref{Scalar-Compatibility} coincide. It is also straightforward to show that our current map $T_\R$ is surjective, and that no condition on the pair $F,G$ needs to be imposed other than \eqref{Scalar-Compatibility}.
It follows that
\begin{equation}
\frac{C_c^\infty(\R)}{\ker T_\R}
\simeq
\mathcal Q_\R,
\label{Scalar-Quotient-Isomorphism}
\end{equation}
which is the non-chiral analogue of
\eqref{Chiral-Smearing-Quotient} for our massless scalar theory.
 Consequently, all the regions considered here have the same
reduced smearing space and generate the same additive scalar
algebra,
\begin{equation}
\mathcal A_\phi(\mathcal D)
=
\mathcal A_\phi(\mathcal C)
=
\mathcal A_\phi(\mathcal B)
=
\mathcal A_\phi(\mathcal T).
\label{Scalar-Equal-Algebras}
\end{equation}
Unlike \eqref{Equal-Chiral-Algebras}, our demonstration is not the result
of chirality but follows from a special property of our massless theory and does not extend to other non-chiral theories. In particular our demonstration does not extend to a massive field. However, this does not exclude equality of the completed local algebras, i.e. when the UV cutoff is removed, when the regions are related by timelike
completion. This will be tested in
Sec.~\ref{sec:massive-scalar-timelike-completion}.

We now define a finite-dimensional truncation directly on the common
quotient space. On the intervals $\{u,v\}\in L$, let
\begin{equation}
h_n(s)
=
e^{i\pi n s/l},
\qquad
|n|\leq n_{\max}\qquad s=u,v.
\end{equation}
A convenient basis of compatible null data is
\begin{equation}
E_0=(h_0,h_0),
\qquad
E_n^{(u)}=(h_n,0),
\qquad
E_n^{(v)}=(0,h_n),
\qquad
0<|n|\leq n_{\max}.
\label{phi-PW-basis}
\end{equation}
Indeed, the non-zero Fourier modes have vanishing integral, while the
coefficients of the two constant modes must agree. 
The truncated space is therefore
\begin{equation}
\mathcal Q_{n_{\max}}
=
\operatorname{span}
\left\{
E_0,E_n^{(u)},E_n^{(v)}:
0<|n|\leq n_{\max}
\right\},
\qquad
\dim\mathcal Q_{n_{\max}}=4n_{\max}+1.
\label{Scalar-Truncated-Quotient}
\end{equation}
To obtain an equality at finite cutoff, we choose the same finite
family of compatible null data for every spacetime realization.
Since the scalar kernels depend only on these marginals, the
corresponding matrix elements coincide entry by entry. Applying
the same reduction to these identical matrices gives equal
generalized spectra. The argument also applies to the disconnected
union, and hence, at fixed $\mu$,
\begin{equation}
I_{\phi,n_{\max}}
(\mathcal X_1:\mathcal Y_2;\mu)
=
I_{\phi,n_{\max}}
(\mathcal D_1:\mathcal D_2;\mu),
\qquad
\mathcal X,\mathcal Y
\in
\left\{
\mathcal D,\mathcal C,\mathcal B,\mathcal T
\right\}.
\label{Scalar-Equal-MI}
\end{equation}

\subsection{Entropy and mutual information for diamond regions}
\label{sec:scalar-diamond-results}

The equality established in
\eqref{Scalar-Equal-MI} makes it sufficient to evaluate the
scalar problem for a single causal diamond and for a union of two
causal diamonds. In this subsection we carry out this calculation in
the quotient-adapted plane-wave basis
\eqref{phi-PW-basis}.

For a single diamond, the symmetry under the exchange
$u\leftrightarrow v$ provides a useful further internal decomposition of the
finite-dimensional problem. For $n\neq0$, define
\begin{equation}
E_n^{(+)}
=
\frac{1}{\sqrt{2}}
\left(
E_n^{(u)}+E_n^{(v)}
\right),
\qquad
E_n^{(-)}
=
\frac{1}{\sqrt{2}}
\left(
E_n^{(u)}-E_n^{(v)}
\right).
\label{Scalar-Symmetric-Basis}
\end{equation}
and assign the common constant mode $E_0$ to the symmetric sector. In this basis and for a single diamond, the matrix elements between the two sectors vanish and the matrix decomposes as
\begin{equation}
\left(
W_{\phi,\mu;n_{\max}}^{\mathcal D},
i\Delta_{\phi;n_{\max}}^{\mathcal D}
\right)
=
\left(
W_{\phi,\mu;n_{\max}}^{\mathcal D,+},
i\Delta_{\phi,n_{\max}}^{\mathcal D,+}
\right)
\oplus
\left(
W_{\phi,\mu;n_{\max}}^{\mathcal D,-},
i\Delta_{\phi,n_{\max}}^{\mathcal D,-}
\right),
\label{Scalar-Diamond-Block-Decomposition}
\end{equation}
where the symmetric block is represented on
$\{E_0,E_n^{(+)}\}$ and the antisymmetric block on
$\{E_n^{(-)}\}$.
This decomposition is closely related to the two families of
Pauli-Jordan eigenfunctions found in \cite{Afshordi:2012ez,Saravani:2013nwa} for this problem,
\begin{align}
f_\kappa(u,v)
&=
e^{-i\kappa u}-e^{-i\kappa v},
&
\kappa
&=\pi n/l,
\qquad n\in\mathbb Z\setminus\{0\},
\label{Scalar-Old-Antisymmetric-Modes}
\\
g_\kappa(u,v)
&=
e^{-i\kappa u}+e^{-i\kappa v}
-2\cos(\kappa l),
&
\tan(\kappa l)
&=
2\kappa l,
\qquad \kappa\neq0.
\label{Scalar-Old-Symmetric-Modes}
\end{align}
Under the reduction map $T_{\mathcal D}$, the first family belongs to
the antisymmetric sector of
\eqref{Scalar-Symmetric-Basis}, while the second belongs to the
symmetric sector. The difference is that
\eqref{Scalar-Old-Antisymmetric-Modes}-\eqref{Scalar-Old-Symmetric-Modes} diagonalize
$i\Delta_\phi$ before the Wightman matrix is constructed. In the
present implementation we retain ordinary Fourier modes in both null
components, remove the elements in the kernel by the SVD procedure and solve the generalized eigenvalue problem directly.
In particular, no roots of the transcendental equation in
\eqref{Scalar-Old-Symmetric-Modes} are required.

The two bases describe the same completed reduced smearing space, but
their finite truncations are not identical. Consequently, equal
nominal values of $n_{\max}$ in the two prescriptions need not give
equal finite matrices or equal convergence rates. Only the continuum
limit and the universal ultraviolet terms should be compared directly.
This is an explicit realization of the distinction between basis
independence and truncation dependence discussed in
Sec.~\ref{Sec:FiniteBasis}.

The common constant mode $E_0$ must be retained when constructing
the kernel matrices. Its omission would restrict both null
marginals to have zero integral and remove the dependence on
$\mu$. Conversely, the absence of a separate constant basis
function in the Pauli--Jordan eigenbasis does not remove this
dependence: the symmetric functions $g_\kappa$ have nonzero
spacetime integrals.

It is also important to distinguish the continuum commutator
from its finite compression. The space $Q_{n_{\max}}$ has odd
dimension, so its antisymmetric commutator form is necessarily
degenerate. For the Fourier truncation above, a null direction is
\begin{equation}
    Z_{n_{\max}}
    =
    E_0+
    \sum_{0<|n|\leq n_{\max}}
    (-1)^n\bigl(E_n^{(u)}+E_n^{(v)}\bigr).
\end{equation}
This is a null direction of the compressed commutator, not a
null smearing of the continuum field, and its Wightman covariance
need not vanish. We construct the full matrices, including $E_0$,
and then select a nondegenerate subspace by the SVD prescription
of Sec.~\ref{Sec:FiniteBasis}. This selection is part of the finite regulator;
it does not identify the finite Fourier truncation with the
finite Pauli--Jordan eigenmode truncation.

We now discuss the expected results for EE and MI in this theory. At fixed infrared parameter $\mu l$, the ultraviolet behavior is
governed by the central charge $c=1$ of the real scalar. Together with
the cutoff identification \eqref{Cutoff-Relation}, this gives
\begin{equation}
S_{\phi,n_{\max}}(\mathcal D;\mu)
=
\frac{1}{3}\log n_{\max}
+
s_{\mathcal D}(\mu l)
+
o(1).
\label{Scalar-Diamond-Entropy-Scaling}
\end{equation}
The coefficient of the logarithm is universal, whereas the constant
$s_{\mathcal D}$ depends on the infrared state and on the precise
ultraviolet truncation. Using the Pauli-Jordan eigenbasis
\eqref{Scalar-Old-Antisymmetric-Modes}-\eqref{Scalar-Old-Symmetric-Modes},
Ref.~\cite{Saravani:2013nwa} report a fitted logarithmic coefficient
$0.33277$ with points below $n_{\max}\leq 2500$, in close agreement with $1/3$. The quotient plane-wave
calculation provides an independent implementation of the same
continuum problem and we get $0.33330$ with points below $n_{\max}\leq 300$. Since the two finite truncations are different,
their non-universal constants should not be expected to agree without
a separate matching of regulators.

We next consider two equal diamonds separated along the spatial
direction. 
On each component we use a copy of the quotient basis
\eqref{phi-PW-basis} adapted to each diamond center. The truncated space of the union is therefore
\begin{equation}
\mathcal Q_{n_{\max}}^{(1)}
\oplus
\mathcal Q_{n_{\max}}^{(2)},
\qquad
\dim
\left(
\mathcal Q_{n_{\max}}^{(1)}
\oplus
\mathcal Q_{n_{\max}}^{(2)}
\right)
=
8n_{\max}+2.
\label{Scalar-Union-Truncated-Space}
\end{equation}
The compatibility condition is imposed separately on the two
components, as required for the additive algebra.
Since the two diamonds are identical, the regulated mutual
information is
\begin{equation}
I_{\phi,n_{\max}}(\eta;\mu l)
=
2S_{\phi,n_{\max}}(\mathcal D;\mu)
-
S_{\phi,n_{\max}}
(\mathcal D_1\cup\mathcal D_2;\mu).
\label{Scalar-Diamond-MI}
\end{equation}
The same value of $n_{\max}$ and the same infrared parameter $\mu$ are
used in all three terms. The commutator has one null
singular direction in each component. Both are removed by the SVD
reduction only after the direct-sum matrices have been assembled.

\begin{figure}[t!]
\centering
\makebox[\textwidth][c]{
\includegraphics[width=1.12\textwidth]{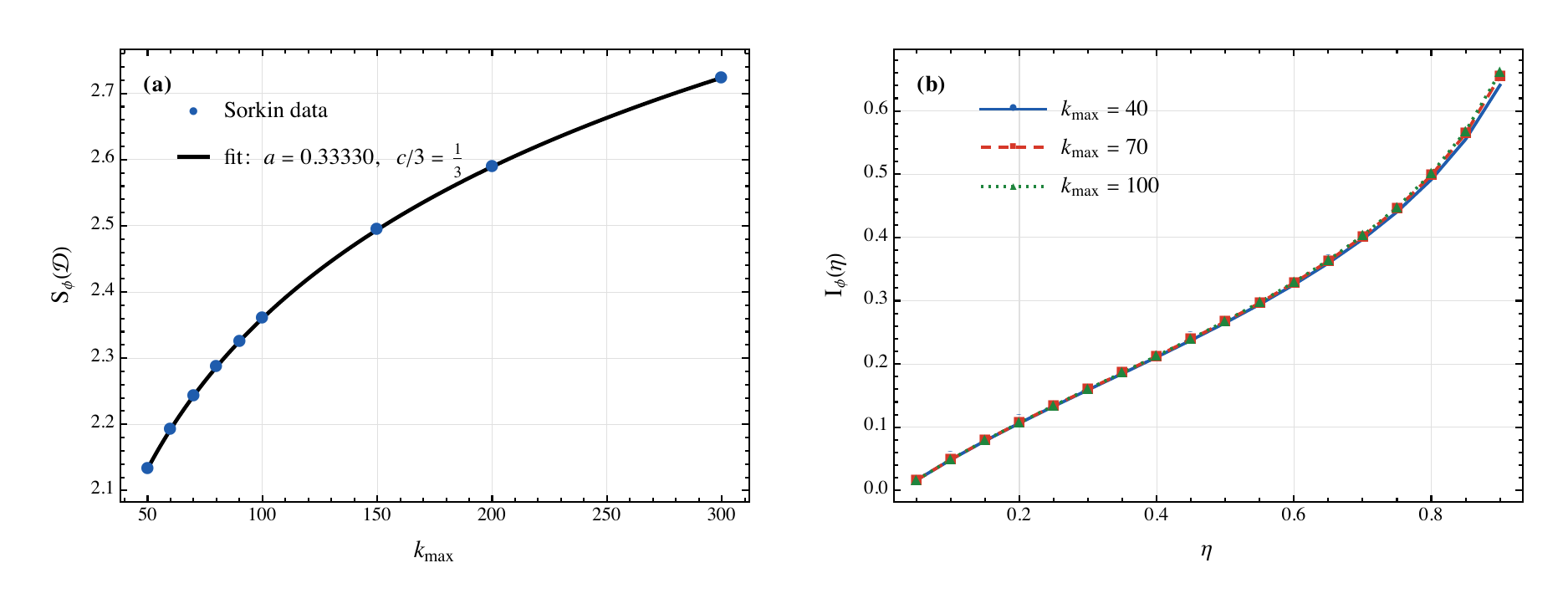}}
\caption{Entropy and mutual information of the infrared-regulated
massless scalar in the quotient-adapted plane-wave truncation, for
$l=1$ and $\mu=1/20$. Panel (a) shows the entropy of a single causal
diamond as a function of $n_{\max}$. The solid black curve is a fit of
the form $a\log n_{\max}+b+c/n_{\max}$, which gives $a=0.33330$, in
agreement with the expected coefficient $c_{\rm CFT}/3=1/3$. Panel
(b) shows the mutual information of two spacelike separated diamonds
as a function of $\eta$ for $n_{\max}=40,70,100$. The curves
converge systematically from below, with the largest finite-cutoff
corrections appearing as $\eta\rightarrow1$.}
\label{fig:scalar-entropy-mi}
\end{figure}

At fixed $\mu l$, the logarithmic ultraviolet
terms in \eqref{Scalar-Diamond-Entropy-Scaling} cancel, and the
continuum mutual information is
\begin{equation}
\mathcal I_\phi(\eta,\mu l)
=
\lim_{n_{\max}\rightarrow\infty}
I_{\phi,n_{\max}}(\eta;\mu l).
\label{Scalar-Diamond-Continuum-MI}
\end{equation}
Unlike the fermionic and current results in Sec.~\ref{Sec:ChiralFields},
the non-compact scalar answer is not a function of $\eta$ alone. It
retains a dependence on the infrared state through the dimensionless
combination $\mu l$.

When \eqref{Scalar-Wightman} is interpreted as the local form of the
Sorkin-Johnston state in a much larger diamond, its validity over the
complete union is controlled by the conditions we described at the beginning of Sec. \ref{sec:scalar-null-reduction}. 
In particular, one could violate the limit by taking $l$ of the size of the large diamond defining the theory $\mu l\sim1$ in which the form \eqref{Scalar-Wightman} would not be a good approximation of the SJ kernel throughout the whole integration domain. Even assuming $\mu l\ll1$, by taking the two diamonds in a MI computation too far away, i.e. $\eta\to0$, we could again be violating the limit in which \eqref{Scalar-Wightman} is a good approximation for the matrix elements involving different regions. In any case, we retain these points in order to exhibit the resulting IR regulator
mismatch, which is analyzed further in Sec.~\ref{sec:scalar-lattice}, and verify that the reduced Wightman matrix remains positive within numerical tolerance throughout the displayed range.

We obverse that the single-diamond entropy converges logarithmically according to
\eqref{Scalar-Diamond-Entropy-Scaling}, while the mutual information
stabilizes substantially faster because its leading ultraviolet terms
cancel. As in the chiral examples, the residual UV cutoff dependence
becomes more pronounced as $\eta\to1$, where the separation
between the diamonds approaches the resolution scale.
Fig.~\ref{fig:scalar-entropy-mi} shows our results for the basis \eqref{Scalar-Symmetric-Basis}. Below, we discuss how the plane-wave cutoff is related quantitatively to a conventional spatial
regulator. We address this by comparison with a massive harmonic-chain
calculation in the next subsection.

\subsection{Comparison with the spatial lattice}
\label{sec:scalar-lattice}

The null-reduction argument of
Sec.~\ref{sec:scalar-null-reduction} establishes the equality between
the different spacetime realizations of the scalar algebra without
requiring a canonical description. Nevertheless, lacking analytical results to compare to, it is useful to
compare the resulting mutual information with an independent
calculation performed on a spatial Cauchy surface. This provides a
numerical benchmark of the plane-wave generalized eigenvalue problem
and, at the same time, makes explicit the relation between its
infrared prescription and a conventional harmonic-chain regulator.
We leave the details of the construction of our spacial lattice for the App. \ref{App:Spatial-Lattice-Details}. The key observation is that we define the adequate spacial lattice theory to compare to \eqref{Scalar-Wightman} by expanding the continuum massive correlator to first order in the mass and matching the coefficient on both expressions. This fixes the mass in the harmonic-chain Hamiltonian in terms of $\mu$ to be $ m= \sqrt{2}e^{-\gamma_{\mathrm E}}\mu$ where $\gamma_{\mathrm E}$ is the Euler-Mascheroni constant. 

The Sorkin calculation is extrapolated to $n_{\max}\to\infty$ independently. We evaluate the plane-wave generalized eigenvalue problem at
\begin{equation}
n_{\max}=40,\ 70,\ 100.
\end{equation}
The cutoff dependence is well described by
\begin{equation}
I_{\phi,n_{\max}}(\eta;\mu)
=
I_{\phi,\infty}(\eta;\mu)
+
\frac{b(\eta)}{n_{\max}},
\label{Scalar-Sorkin-Extrapolation}
\end{equation}
which we use as the extrapolation ansatz. As a stability check, we
compare the fit using the three cutoffs with the extrapolation obtained
from the two largest values. Their difference remains below
$7.4\times10^{-4}$ for the range of cross-ratios considered here.

\begin{figure}[t!]
\centering
    \makebox[\textwidth][c]{\includegraphics[width=1.12\textwidth]{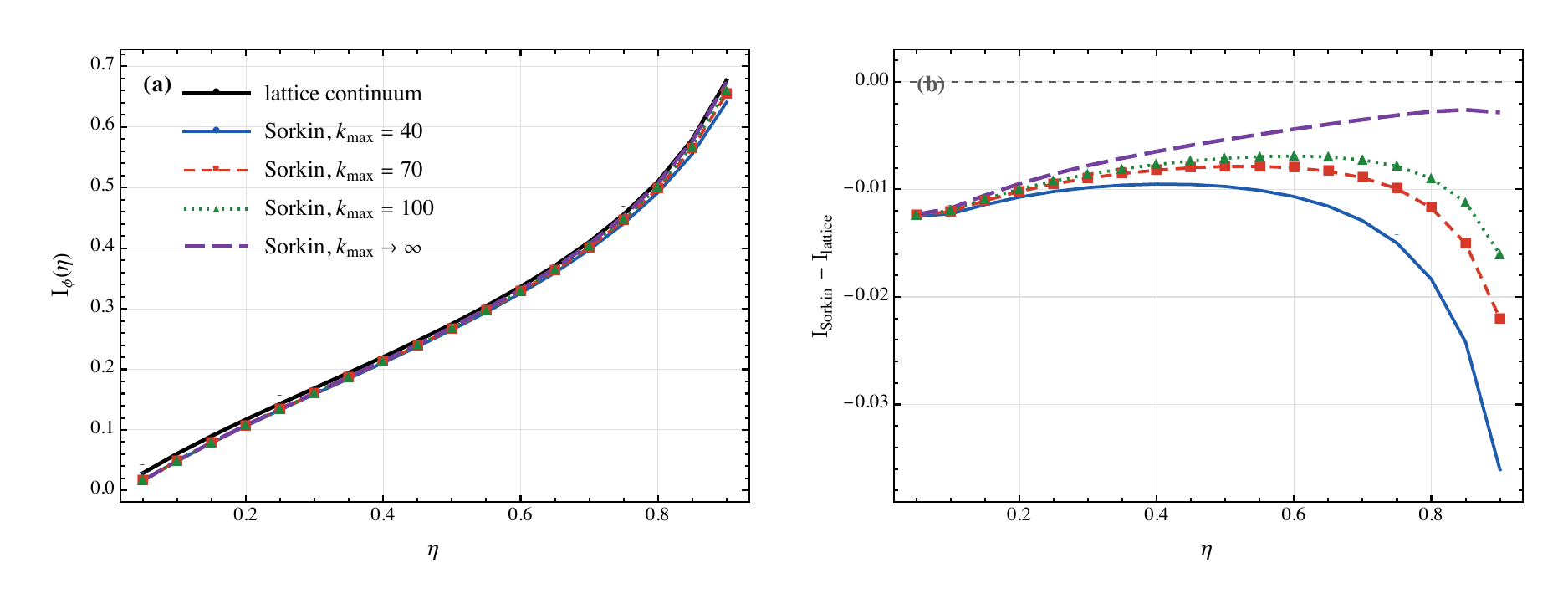}}
\caption{Mutual information of two intervals for the
infrared-regulated massless scalar. Panel (a) compares the
continuum-extrapolated harmonic-chain result with the Sorkin
generalized eigenvalue problem at
$n_{\max}=40,70,100$. The dashed purple curve is obtained by
extrapolating the Sorkin results linearly in $1/n_{\max}$ according to
\eqref{Scalar-Sorkin-Extrapolation}. Panel (b) shows the signed difference between the Sorkin curves
and the lattice extrapolation.
The finite-$n_{\max}$ results converge systematically from below.
The remaining discrepancy at small $\eta$ is insensitive to the
plane-wave cutoff and originates from the fact that the massive
lattice correlator agrees with the logarithmic kernel only in the
short-distance regime.}
\label{fig:scalar-sorkin-lattice}
\end{figure}

Using this convention, Figure~\ref{fig:scalar-sorkin-lattice} compares the two calculations
for $l=1$ and $m=1/20$, 
where the finite-cutoff behavior and the regulator-matching effect can be
clearly separated. 
As $\eta\to1$, the intervals approach each other and increasingly
short-distance modes contribute to the mutual information. The
difference between the three finite-$n_{\max}$ curves therefore grows
in this regime. Nevertheless, the extrapolated Sorkin result is close
to the lattice value. For example,
\begin{equation}
I_{\phi,\infty}(0.9;\mu)
\simeq
0.6739,
\qquad
I_{\mathrm{lat}}(0.9)
\simeq
0.6768\,,
\end{equation}
\begin{equation}
I_{\phi,\infty}(0.5;\mu)
\simeq
0.2696,
\qquad
I_{\mathrm{lat}}(0.5)
\simeq
0.2749,
\end{equation}
corresponding to a difference of approximately $0.4\%$ and $2\%$ respectively. At small
cross-ratio, instead, the three finite-$n_{\max}$ Sorkin curves have
already converged but remain below the lattice result, e.g. 
\begin{equation}
I_{\phi,\infty}(0.05;\mu)
\simeq
0.01647,
\qquad
I_{\mathrm{lat}}(0.05)
\simeq
0.02880.
\end{equation}
Although the absolute difference remains small and of the same order as for $\eta \sim 0.5$, its relative value is
large because the mutual information itself is small in this limit.
This remaining difference is consistent with the range of validity of
the prescription we gave in comparing the models. In expanding the continuum correlator for small mass, we  assumed that no distance involved in the computation was bigger than the mass. However, for sufficiently small $\eta$ the systems are so far away that they begin competing with the IR regulator and the log approximations fails. We have already discussed a similar effect directly in \eqref{Scalar-Wightman} to find that said equation is not to be trusted for small $\eta$ for analogous reasons.
We stress that increasing $n_{\max}$ cannot remove this difference, since it is an infrared
matching effect rather than a UV truncation error.

The comparison should therefore not be interpreted as an exact
regulator-independent equality at fixed $\mu$. Instead, it provides
an independent canonical benchmark of the quotient-adapted Sorkin
calculation in the regime where the two infrared prescriptions
describe the same local correlations. In particular, the agreement
towards $\eta \sim 1$, together with the independent convergence of
both regulators, supports the plane-wave construction of
Sec.~\ref{sec:scalar-diamond-results}. Finally, by
\eqref{Scalar-Equal-MI}, the same Sorkin curve applies to all the
spacetime realizations of the same algebra considered in this work.

\section{The massive scalar}
\label{sec:massive-scalar-timelike-completion}

We now study a massive scalar field in the Minkowski vacuum, which is a non-chiral UV complete QFT example. Consequently, causality and the timelike tube theorem can be used to prove \eqref{Completion-MI-Prediction} but we only expect a match in the $n_{\max}\to\infty$ limit, unlike in our previous examples where we were able to prove the matching at each fixed $n_{\max}$. We could in principle consider all possible combinations of regions in \eqref{Completion-MI-Prediction} but we find that the least trivial one is the one relating the MI between two timelike lenses and two diamonds, 
so we focus on testing this result.

\subsection{Local algebras and spacetime truncations}
\label{sec:massive-scalar-algebras}

We work directly with the massive Minkowski Wightman function. A useful
mass-shell representation is
\begin{equation}
W_m(x,x')
=
\int_{-\infty}^{\infty}
\frac{d\theta}{4\pi}
\exp\!\left[
-i\frac{m}{\sqrt{2}}e^\theta(u-u'-i0^+)
-i\frac{m}{\sqrt{2}}e^{-\theta}(v-v'-i0^+)
\right],
\label{Massive-Wightman-Rapidity}
\end{equation}
and the commutator matrix is obtained from the antisymmetric part.

The equations of motion for the massive scalar can be written as
\begin{equation}
P_m=\Box-m^2=-2\partial_u\partial_v-m^2
\label{Massive-KG-Operator}
\end{equation}
The field equation implies that
two test functions which differ by $P_m h$, with
$h\in C_c^\infty(\R)$, define the same smeared field. The corresponding
algebraic test-function space can be written as
\begin{equation}
\mathcal E_m(\R)
=
\frac{C_c^\infty(\R)}{P_m C_c^\infty(\R)}.
\label{Massive-Test-Function-Quotient}
\end{equation}
We find convenient to use the same family of massive plane waves for all regions,
\begin{equation}
f_{n,\pm}(t,x)
=
e^{\mp i\omega_n t+ip_nx},
\qquad
p_n=\frac{\pi n}{\sqrt{2}l},
\qquad
\omega_n=\sqrt{p_n^2+m^2},
\qquad
|n|\leq n_{\max}.
\label{Massive-Plane-Wave-Basis}
\end{equation}
The dimension of the truncated space is therefore $4n_{\max}+2$ per
connected component. 

For the theory at hand, find convenient to define a
one-parameter family of straight-sided timelike lenses 
\begin{equation}
\mathcal R_\rho
=
\left\{
(t,x):
|t|
+
\frac{|x|}{\rho}
<\sqrt{2}l
\right\},
\qquad
\mathcal R_1=\mathcal D.
\qquad
E(\mathcal R_\rho)
=
\mathcal D
\qquad
0<\rho\leq1.
\label{Straight-Lens-Definition}
\end{equation}
At fixed $t$, the spatial width of $\mathcal R_\rho$ is a fraction
$\rho$ of the width of the reference diamond. 

The timelike tube theorem then gives
\begin{equation}
\mathcal A_m(\mathcal R_\rho)
=
\mathcal A_m(\mathcal D),
\qquad
0<\rho\leq1,
\label{Massive-Timelike-Algebra-Equality}
\end{equation}
where the equality is understood for the completed local operator
algebras \cite{Borchers,Araki,Witten:2023aro,Strohmaier:2023ttt}.
This implies that the individual smeared operators in the diamond  will not in general be
finite linear combinations of operators smeared in the lens and that their
reconstruction may limits of increasingly complicated
combinations of lens-supported operators, i.e. the truncated spaces of both computations will in general generate different algebras. In other words, unlike in our previous examples, the mutual information between the diamonds and between the lenses is not expected to agree until $n_{\max}\to\infty$ and may approach the same continuum results at different rates. In any case, for the sake of the comparison, we choose to keep same number of modes for the two computations. We then apply the same SVD reduction and generalized eigenvalue problem as in
Sec.~\ref{Sec:FiniteBasis}.

\subsection{Mutual information and timelike-tube convergence}
\label{sec:massive-scalar-results}

We consider two diamonds and two equal, spacelike separated copies of
$\mathcal R_\rho$ and we use the cross-ratio
$\eta$ defined in \eqref{Cross-Ratio}. 
Since
\eqref{Massive-Timelike-Algebra-Equality} holds for each connected
component, the continuum mutual information must satisfy
\begin{equation}
I_m(\mathcal R_{\rho,1}:\mathcal R_{\rho,2})
=
I_m(\mathcal D_1:\mathcal D_2)\qquad
0<\rho\leq1.
\label{Massive-Timelike-MI-Equality}
\end{equation}
At finite cutoff, however, no such equality is expected. The numerical
prediction is instead
\begin{equation}
\lim_{n_{\max}\rightarrow\infty}
\left[
I^{(m)}_{\rho,n_{\max}}(\eta)
-
I^{(m)}_{1,n_{\max}}(\eta)
\right]
=
0.\qquad
0<\rho\leq1.
\label{Massive-Timelike-MI-Convergence}
\end{equation}
We test \eqref{Massive-Timelike-MI-Convergence} for the representative
choice
\begin{equation}
l=1,
\qquad
m=\frac{3}{10},
\qquad
\rho=\frac{1}{2}.
\label{Massive-Numerical-Parameters}
\end{equation}
The spatial interval whose causal development is the reference diamond
has length
\begin{equation}
L=2\sqrt{2}l,
\qquad
mL=\frac{3\sqrt{2}}{5}\simeq0.85.
\end{equation}
Thus the lens has half the spatial width of the diamond at every time,
while the mass is neither in the nearly massless regime nor large
enough to suppress the mutual information trivially. 

\begin{figure}[t!]
\centering
 \makebox[\textwidth][c]{
\includegraphics[width=1.12\textwidth]{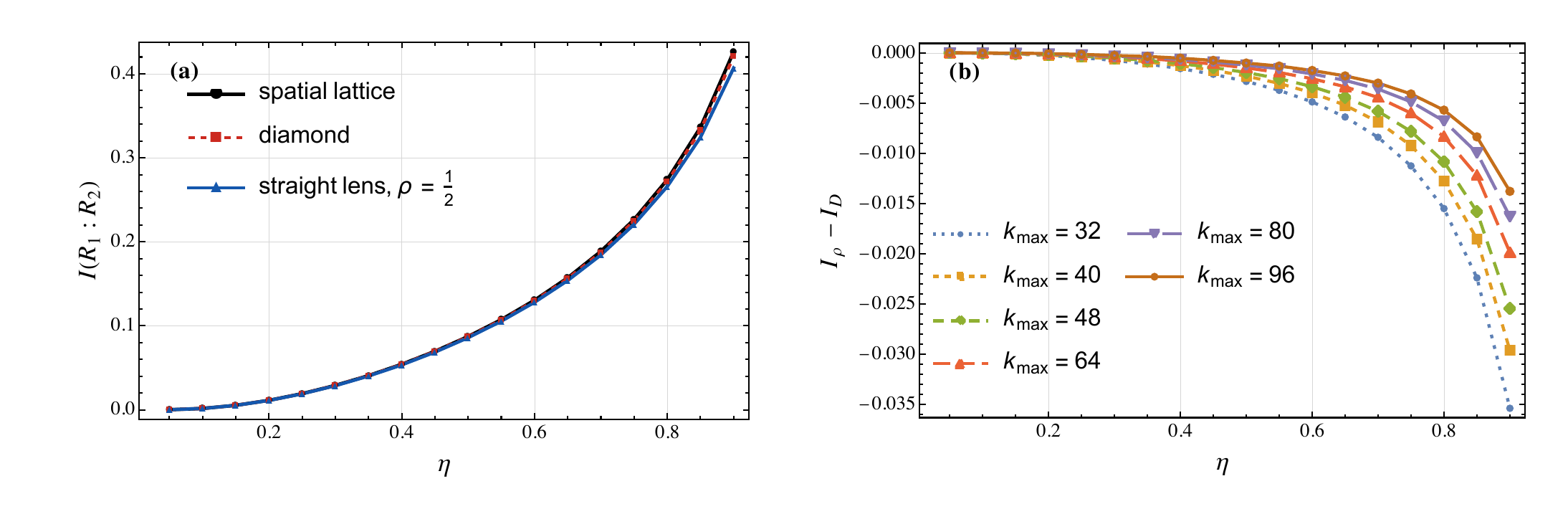}
}
\caption{Mutual information of the massive scalar for two spacelike
separated regions, with $l=1$, $m=3/10$ and $\rho=1/2$. Panel (a)
compares the continuum-extrapolated spatial-lattice result with the
spacetime generalized eigenvalue problem for two diamonds and two
straight-sided timelike lenses at $n_{\max}=96$. Panel (b) shows
$I^{(m)}_{1/2,n_{\max}}-I^{(m)}_{1,n_{\max}}$ for
$n_{\max}=32,40,48,64,80,96$. The difference decreases systematically as
the cutoff is increased. The slower convergence as $\eta\to1$ reflects
the increasing sensitivity to short-distance modes when the two
associated diamonds approach each other.}
\label{fig:massive-scalar-lens-mi}
\end{figure}

We compare our results with a spatial
lattice obtained from the harmonic-chain construction of
App.~\ref{App:Spatial-Lattice-Details}, using the same mass $m=3/10$ and extrapolating
the lattice spacing to zero. 
Figure~\ref{fig:massive-scalar-lens-mi} summarizes the result. Panel
(a) compares the continuum-extrapolated spatial lattice with the
diamond and straight-lens spacetime calculations at $n_{\max}=96$.
Panel (b) shows the difference
$I^{(m)}_{1/2,n_{\max}}-I^{(m)}_{1,n_{\max}}$ for several values of the
plane-wave cutoff.

The diamond curve approaches the independent lattice result over the
complete range shown, with a relative difference below $1.4\%$ at
$n_{\max}=96$. The straight-edged lens also moves systematically
towards the diamond as the cutoff is increased, but converges more slowly in agreement with our comment around footnote \ref{Fewster}. At $n_{\max}=96$, the
relative difference between the two spacetime calculations is about
$1.2\%$ at $\eta=0.5$ and $3.3\%$ at $\eta=0.9$. 
The convergence is slower near $\eta=1$, as expected, since the
separation between the two diamonds becomes comparable with the
UV resolution scale. Reconstructing the diamond algebra from
the lens also requires increasingly oscillatory combinations
of lens-supported modes, making this geometry more sensitive to the
finite truncation.
The residual difference is well above the numerical errors. At the
most demanding point displayed, $n_{\max}=64$ and $\eta=0.9$, the
off-diagonal commutator block gives a microcausality residual of order
$10^{-12}$, while the generalized-eigenvalue pairing residual is
approximately $3.2\times10^{-6}$. We interpret the separation between the lens and
diamond curves as a finite-cutoff effect
rather than a loss of numerical precision.

The numerical convergence in Fig.~\ref{fig:massive-scalar-lens-mi}, together with the
independent spatial-lattice benchmark, provides a direct
test of the timelike tube theorem within the spacetime entropy
prescription.

\section{Discussion}
\label{sec:discussion}

We have studied the covariant prescription given in \cite{Sorkin2012,Saravani:2013nwa} to obtain entanglement entropy from spacetime correlators in a number of examples involving free theories and directly in the spacetime continuum. 
Our main contribution is to develop a number of techniques within this prescription to make it more straightforward to apply to examples of deep interest to the community. In particular, we pointed out and proved in several examples that whilst most uses in the literature of this prescription attempt an analytic diagonalization of the reduced Pauli Jordan operator $i\Delta_\R$ within the region\footnote{It is fair to say that these works are more motivated in exploring a local definition for a vacuum of a theory in the sense of the Sorkin-Jonston prescription \cite{Sorkin:2011pn,Johnston:2009fr}, for which an $i\Delta_\R$ diagonalization cannot be avoided.}, which has proven to be a formidable task even for a free massive scalar in a diamond \cite{Mathur:2019yvl}, that such an analytic decomposition can be replaced in numeric calculations with a SVD process that effectively removes the elements in the kernel starting from a standard plane wave basis. Our results reproduce the expected continuum behavior and agree
with the available analytical and spatial-lattice benchmarks
in their common regime of validity.

Our main interest for this prescription in this work is that it allows to directly test several properties of covariance in entanglement measures for relativistic quantum field theories. A familiar example is the comparison between the entanglement entropy obtained from two different Cauchy surfaces of the same subsystems in theories with a local stress tensor. In such examples, the equations of motion allow to explicitly map the information in one surface to another. Put in algebraic terms, it can be said that the operator algebra generated by the two Cauchy data are the same, and since entanglement measures are defined in terms of operator algebras, it must follow that both yield identical results. 
A less familiar example that we were interested in exploring was the timelike tube theorem that state that, regarless of the existence of a local stress tensor in the theory, the algebra generated by a timelike tube (or lens) is the same as the one generated by its envelope, see Fig. \ref{fig:spacetime-regions-2}. We found no test of this standard result obtained in the context of Algebraic QFT and we aimed to fill this gap by developing a numerical application of the covariant prescription in \cite{Sorkin2012,Saravani:2013nwa}, which is the natural environment for this test. We know of no other prescription in the literature that is capable of such a test.

The description in terms of smeared fields also makes the
algebraic equivalences explicit. For chiral fields, the relevant
data are smearings on a single null projection; for the massless
scalar, they are pairs of compatible null marginals. Choosing
the same finite family of these data for different spacetime
realizations makes their kernel matrices identical at every
cutoff. This common null-data construction is special to the
massless examples considered here and is not available for
our massive plane-wave truncations.

We have chosen the examples in this work to highlight a number of nice properties of the formalism. 
For the Weyl fermion, we chose the same basis for different spacetime regions with the same null projection and found perfect agreement between all results as the UV cut-off is removed. In particular, we found that in some sense a timelike tube realizes the operator algebra of a segment more effectively than the more symmetrical diamond shaped regions, see Fig.~\ref{fig:Weyl-MI}. For the chiral current, in contrast, we exploited the smeared field space and showed that by choosing a particular basis we can match the operator algebras generated by any spacetime region to the one generated on the null line at each value of finite cutoff. This implies that, in that basis and to each value of the cut-off, all regions generate the same operator algebra as the null projection and no further comparison between different regions was required. We were able to extend this formalism to the IR regulated massless scalar. Hence, in this example, where an explicit non-trivial diagonalization of $i\Delta^\R$ for the diamond is known, we focused instead on showing that one can reproduce standard lattice results for the mutual information using a naive plane-wave basis and then reduce to the kernel via a standard SVD method. The massive scalar provides a final test of the timelike tube theorem in this work, since no null projection is available and the algebras generated by e.g. the diamond and a timelike lens become the same only when the cutoff is removed. All our results are consistent with the results of the timelike tube theorem.

Having already established this method as an relevant alternative on its own for the benchmark free theories examples, our immediate goal is to apply the spacetime prescription to generalized free
fields \cite{Greenberg:1961mr,Duetsch:2002hc,Yngvason:1994nk}. These theories are Gaussian and are defined directly by their spacetime two-point functions, which can be written in terms of a
Källén-Lehmann spectral density as
\begin{align}
W_{\mathrm{GFF}}(x,y)
&=
\int_0^\infty dM^2\,
\varrho(M^2)\,W_M(x,y),
\nonumber\\
i\Delta_{\mathrm{GFF}}(x,y)
&=
\int_0^\infty dM^2\,
\varrho(M^2)\,i\Delta_M(x,y).
\label{GFF-Spectral-Representation}
\end{align}
The finite-basis matrices entering the spacetime entropy prescription
can consequently be constructed directly from the corresponding
spectral superpositions and no canonical variables or reduced density
matrix on a preferred Cauchy slice are required. This is the main advantage of the current spacetime formalism since for a general nontrivial spectral density there is no local equation of motion
that reduces the spacetime smearing space, and the time-slice property
can fail. Interestingly, one should also find that the timelike tube theorem, which does not rely on a local equation of motion, still holds, i.e. that a timelike lens produces the same MI than the full diamond algebra. Generalized free fields parametrized by different $\varrho(M^2)$ therefore provide a natural setting
in which the spacetime formulation is not merely an alternative
description of a standard calculation but rather the only approach available in the literature. 
Studying their mutual
information for causally and timelike related spacetime presentations
should reveal how the reduced algebra and its completion depend on the
spectral density, and provides the immediate continuation of the
present work.

It should also be mentioned that, as was pointed out in e.g. \cite{Duetsch:2002hc}, for the particular case of $\varrho(M^2)= M^{2\nu}$ with $d$ the spacetime dimensions, the resulting GFF correlators have conformal symmetry and a holographic dual in the sense of AdS/CFT \cite{Aharony:1999ti} can be build by considering the boundary correlators of a single massive scalar of mass $m^2$ in a pure AdS background of radius $R_{AdS}$ such that $\nu=\sqrt{\frac{d^2}{4}+ m^2R_{AdS}^2}$. Explorations of this holographic GFF were made on the AdS side e.g. \cite{Benedetti:2022aiw,CesarThesis}, but we don't know of any available exploration of such theories directly on the QFT side. We will close this gap in an upcoming work.

A more speculative, but possible line of work that follows from our results is the exploration of entanglement measures for systems which are not necessarily spacelike separated\footnote{We thank Cesar Agón for pointing out this possibility.}. We mentioned this possibility around eq. \eqref{dubious} but we are cautious to denote this EE combination as a true mutual information, since a proper statistical interpretation of the quantity requires commuting degrees of freedom between the systems $\R_1$ and $\R_2$ which is not met for general timelike separated systems. In any case, this type of questions have appeared recently in many works resorting in general to the complexification of standard methods or using holography \cite{published_papers/18375345, Doi2023Timelike, Anegawa2024Black}. We have carried out preliminary explorations of two diamonds on top of each other using \eqref{dubious} versus two spacelike separated diamonds as in the main text and found that the results for the timelike separated diamonds are qualitatively similar to the MI but lie above the proper mutual information curves for all $\eta$, consistent with the intuitive idea that timelike separated spacetime regions are allowed to share more information. 

As a concluding remark, our results support the interpretation of the
generalized spectral prescription as an information-theoretic
construction associated with local operator algebras rather than with
a particular spatial slice or spacetime representative. The examples
studied here show explicitly how this algebraic content may be visible
already at finite cutoff and how they emerge as the regulator
is removed.

\section*{Acknowledgments}
We thank Horacio Casini, Christopher J. Fewster and Cesar A. Agón for helpful discussions. RA and PJM were supported by CONICET and UNLP, Argentina. M.H. is supported by CONICET, CNEA, and Instituto Balseiro, Universidad Nacional de Cuyo. M.H. would like to acknowledge also support from the ICTP through the Associates Programme (2026-2030). The authors are indebted to the Simons Foundation targeted grant to institutions that funds the Simons Balseiro Theoretical Physics Initiative, as an important part of the present work was discussed at the 'Simons-Balseiro School and Workshop on Advanced Aspects of QFT' at Bariloche, Argentina.

\newpage

\appendix

\section{Details on the spatial lattice}
\label{App:Spatial-Lattice-Details}

Our spatial lattice model is defined as follows. We discretize a massive scalar on a periodic spatial lattice with $N$ sites
and lattice spacing $a$. Up to an overall normalization which does not
affect the ground state, the harmonic-chain Hamiltonian is
\begin{equation}
H_{\mathrm{lat}}
=
\frac{1}{2}
\sum_{j=0}^{N-1}
\left[
p_j^2
+
(q_{j+1}-q_j)^2
+
(ma)^2q_j^2
\right],
\qquad
[q_j,p_k]=i\,\delta_{jk},
\label{Scalar-Lattice-Hamiltonian}
\end{equation}
with periodic boundary conditions $q_N=q_0$. The normal-mode
frequencies are
\begin{equation}
\omega_n
=
\sqrt{
(ma)^2
+
4\sin^2\left(\frac{\vartheta_n}{2}\right)
},
\qquad
\vartheta_n=\frac{2\pi n}{N}
\label{Scalar-Lattice-Dispersion}
\end{equation}
such that the ground-state covariance matrices are
\begin{align}
X_{jk}
&=
\langle q_jq_k\rangle
=
\frac{1}{N}
\sum_{n=0}^{N-1}
\frac{
e^{i\vartheta_n(j-k)}
}{
2\omega_n
},
\label{Scalar-Lattice-X}
\\
P_{jk}
&=
\langle p_jp_k\rangle
=
\frac{1}{N}
\sum_{n=0}^{N-1}
\frac{\omega_n}{2}
e^{i\vartheta_n(j-k)}.
\label{Scalar-Lattice-P}
\end{align}
For a set of lattice sites $A$, let $X_A$ and $P_A$ denote the
corresponding restricted matrices. If $\nu_\alpha$ are the positive
symplectic eigenvalues,
\begin{equation}
\nu_\alpha
=
\sqrt{
\operatorname{eig}_\alpha(X_AP_A)
},
\qquad
\nu_\alpha\geq\frac{1}{2},
\end{equation}
the entropy is
\begin{equation}
S_{\mathrm{lat}}(A)
=
\sum_\alpha
\left[
\left(\nu_\alpha+\frac{1}{2}\right)
\log\left(\nu_\alpha+\frac{1}{2}\right)
-
\left(\nu_\alpha-\frac{1}{2}\right)
\log\left(\nu_\alpha-\frac{1}{2}\right)
\right].
\label{Scalar-Lattice-Entropy}
\end{equation}
This is the standard correlation-matrix construction for bosonic
Gaussian systems
\cite{Peschel:2002yqj,PeschelEisler:2009,Audenaert:2002}.
The lattice mutual information is then computed as
\begin{equation}
I_{\mathrm{lat}}(A_1:A_2)
=
S_{\mathrm{lat}}(A_1)
+
S_{\mathrm{lat}}(A_2)
-
S_{\mathrm{lat}}(A_1\cup A_2).
\label{Scalar-Lattice-MI}
\end{equation}

In comparing with the massless scalar of Sec.~\ref{sec:massless-scalar} the theories are compared as follows. We begin by noticing that at $v=-u$, or $t=0$ we can write \eqref{Scalar-Wightman} as, see \eqref{Lightcone-Coordinates}
\begin{equation}
W_{\phi,\mu}(0,x;0,x')
=
-\frac{1}{2\pi}
\log\left(\frac{\mu|x-x'|}{\sqrt{2}}\right).
\label{Scalar-Equal-Time-Convention}
\end{equation}

Now, the continuum equal-time correlator of a scalar of mass $m$ is
\begin{equation}
X_m(r)
=
\frac{1}{2\pi}K_0(m|r|),
\end{equation}
where $K_0$ is the modified Bessel function. At short distances,
\begin{equation}
K_0(m|r|)
=
-\log\left(\frac{m|r|}{2}\right)
-\gamma_{\mathrm E}
+
\mathcal O\left(
m^2r^2\log(m|r|)
\right).
\end{equation}
Matching the leading short-distance expressions, including
their additive constants, gives
\begin{equation}
m
=
\sqrt{2}e^{-\gamma_{\mathrm E}}\mu.
\label{Scalar-IR-Matching}
\end{equation}
where $\gamma_{\mathrm E}$ is the Euler-Mascheroni constant. This identification matches the correlators in the regime
$m|r|\ll 1$; it does not identify the two theories at arbitrary
separations. 

We now discuss the definition of the cross-ratio $\eta$ in \eqref{Cross-Ratio} on the spatial lattice. The spatial section of a diamond at $t=0$ is an interval of length
\begin{equation}
L_A=2\sqrt{2}\,l.
\end{equation}
For two equal intervals whose centers are separated by $d_x$, the
cross-ratio used in the spacetime calculation is
\begin{equation}
\eta
=
\frac{L_A^2}{d_x^2}.
\label{Scalar-Lattice-Cross-Ratio}
\end{equation}
An interval containing $n_A$ lattice sites therefore has
\begin{equation}
a=\frac{L_A}{n_A},
\qquad
\frac{d_x}{a}
=
\frac{n_A}{\sqrt{\eta}}.
\label{Scalar-Lattice-Separation}
\end{equation}
Since the second quantity need not be an integer, at each resolution
we compute the mutual information at the two nearest integer
separations and interpolate linearly to the desired value of $\eta$.

We use
\begin{equation}
n_A
\in
\left\{
40,60,80,120,160
\right\}.
\end{equation}
For every resolution, the circumference of the periodic chain is
chosen such that $mL_{\mathrm{box}}\geq18$, while the two intervals
occupy only a small fraction of the complete chain. Finite-volume
effects are consequently exponentially suppressed. At fixed $\eta$,
the central continuum extrapolation is obtained from the four finest
resolutions using
\begin{equation}
I_{\mathrm{lat}}(\eta,a)
=
I_{\mathrm{lat}}(\eta)
+
c_2(\eta)
\left(\frac{a}{l}\right)^2.
\label{Scalar-Lattice-Extrapolation}
\end{equation}
The stability of the extrapolation is estimated by comparing this
result with a fit using all resolutions and with a fit including a
term linear in $a/l$. 
The resulting spread is at most of order $10^{-3}$ over the
range of cross-ratios shown in the main text.

\newpage

\bibliography{SorkinRefs}{}
\bibliographystyle{ytphys} 

\end{document}